\documentclass[journal]{IEEEtran}
\usepackage{xurl}
\usepackage{dblfloatfix}
\usepackage{tabularx}
\usepackage{wrapfig}
\usepackage{amsmath}
\usepackage{color}
\usepackage{amsmath,amsfonts,amssymb,amscd}
\usepackage{cite}
\usepackage{booktabs}
\usepackage{paralist}
\usepackage{enumitem}
\usepackage[export]{adjustbox}
\usepackage{gensymb}
\usepackage{steinmetz}
\usepackage{mathtools}
\usepackage{siunitx}
\usepackage{hyperref}
\usepackage{cuted}

\usepackage{amsmath,bm}

\usepackage{amssymb}
\usepackage{nomencl}
\makenomenclature

\renewcommand{\nomgroup}[1]{%
  \ifstrequal{#1}{A}{\item[\textit{Acronyms}]}{%
  \ifstrequal{#1}{I}{\vspace{10pt}\item[\textit{Indices and sets}]}{%
  \ifstrequal{#1}{P}{\vspace{10pt}\item[\textit{Parameters}]}{%
  \ifstrequal{#1}{V}{\vspace{10pt}\item[\textit{Variables}]}{}}}}
}

\usepackage[noabbrev]{cleveref}

\usepackage{mathrsfs}
\usepackage{multirow}
\usepackage{flushend}
\usepackage{graphicx}
\usepackage{algorithm2e}

\usepackage{pifont}
\newcommand{\xmark}{\ding{55}}
\newcommand{\cmark}{\ding{51}}

\usepackage{multicol}

\begin{document}
\thispagestyle{empty}
\onecolumn
\begin{quote}
    This paper has been \textcolor{blue}{accepted} for publication in \textcolor{blue}{IEEE Transactions on Smart Grid}
\end{quote}

\vspace{1cm}
\begin{quote}
     doi: 10.1109/TSG.2023.3339707
\end{quote}
\vspace{1cm}
\begin{quote}
    Content is final as presented here, with the exception of pagination.
\end{quote}

\vspace{1cm}
\begin{quote}
    IEEE Copyright Notice: 

\vspace{0.25cm}
\noindent
\copyright 2023 IEEE. Personal use of this material is permitted.  Permission from IEEE must be obtained for all other uses, in any current or future media, including reprinting/republishing this material for advertising or promotional purposes, creating new collective works, for resale or redistribution to servers or lists, or reuse of any copyrighted component of this work in other works.
\end{quote}

\newpage
\twocolumn
\title{Delay-Aware Semantic Sampling in Power Electronic Systems\\
\thanks{This work is supported by the Nordic Energy Research programme via Next-uGrid project n. 117766.

Kirti Gupta and Bijaya Ketan Panigrahi are with the Department of Electrical Engineering, Indian Institute of Technology Delhi, New Delhi 110016, India (e-mail: \{\texttt{Kirti.Gupta, Bijaya.Ketan.Panigrahi}\}@ee.iitd.ac.in).}
\thanks{Subham Sahoo is with the Department of Energy, Aalborg University, 9220 Aalborg, Denmark (e-mail: sssa@energy.aau.dk).}
}

\author{Kirti~Gupta,
        Subham~Sahoo,~\IEEEmembership{Senior Member,~IEEE},
        and Bijaya~Ketan~Panigrahi,~\IEEEmembership{Fellow,~IEEE}
        }

\maketitle
\setcounter{page}{1}
\begin{abstract}
In power electronic systems (PES), attacks on data availability such as latency attacks, data dropouts, and time-synchronization attacks (TSAs) continue to pose significant threats to both the communication network and the control system performance. As per the conventional norms of communication engineering, PES still rely on time synchronized sampling, which translates every received message with equal importance. In this paper, we go beyond event-triggered sampling/estimation to integrate \textit{semantic} principles into the sampling process for each distributed energy resource (DER), which not only compensates for delayed communicated signals by reconstruction of a new signal from the inner control layer dynamics, but also evaluates the reconstruction stage using key semantic requirements, namely \texttt{Freshness}, \texttt{Relevance} and \texttt{Priority} for good dynamic performance. As a result, the sparsity provided by event-driven sampling of internal control loop dynamics translates as \textit{semantics} in PES. The proposed scheme has been extensively tested and validated on a modified IEEE 37-bus AC distribution system, under many operating conditions and noisy environment in OPAL-RT environment to establish its robustness, model-free design ability and adaptive behavior to dynamic cyber graph topologies. 

\end{abstract}


\IEEEpeerreviewmaketitle

\printnomenclature[1in]  

\nomenclature[I]{$j$, $m$}{Index of DERs in PES}
\nomenclature[I]{$\mathrm{N}_{j}$}{The set of neighbouring DERs to $j^{th}$ DER}
\nomenclature[P]{$\mathrm{a}_{jm}$}{Communication weight}
\nomenclature[P]{$\mathrm{\sigma}_m$}{Information received from $m^{th}$ DER}
\nomenclature[P]{$\Delta{\omega \mathrm{c}}_{j}$ and $\Delta \mathrm{Vc}_{j}$}{Frequency and voltage correction term from secondary controller of $j^{th}$ DER}
\nomenclature[P]{$\mathrm{m}^{\mathrm{p}}_{j}$ and $\mathrm{n}^{\mathrm{q}}_{j}$}{Active and reactive power droop coefficient of $j^{th}$ DER}
\nomenclature[P]{$\mathrm{g}_j$}{Convergence parameter of $j^{th}$ DER}
\nomenclature[P]{$\mathrm{\tau}_m$}{Time delay from $m^{th}$ DER}
\nomenclature[P]{$\alpha$}{Alpha, a tunable parameter}
\nomenclature[P]{$D$}{Downsampling factor}
\nomenclature[P]{$\mathrm{F}$ and $\mathrm{R}$}{Freshness and relevance}
\nomenclature[P]{$\mathrm{u(t)}$}{Timestamp of the latest packet received at destination by time $\mathrm{t}$}

\nomenclature[V]{$\textbf{e}_j^{\textbf{dqVC}}$}{Vector of error signals provided to local voltage controller of $j^{th}$ DER}
\nomenclature[V]{$\mathrm{k_1}$ and $\mathrm{k_2}$}{Tunable gains}
\nomenclature[V]{$\textbf{u}_j^{\textbf{pq}}$}{Vector of local control inputs to secondary controller of $j^{th}$ DER}
\nomenclature[V]{$\textbf{e}_j^{\textbf{dqD}}$}{Vector of downsampled signal of $j^{th}$ DER}
\nomenclature[V]{$\mathrm{t}_a$}{Triggering moment}
\nomenclature[V]{$\textbf{e}_j$}{Vector of error of $j^{th}$ DER, fed to the prediction policy}
\nomenclature[V]{$\textbf{e}_j^{\textbf{R}}$}{Vector of reconstructed signals of $j^{th}$ DER}
\nomenclature[V]{$\textbf{u}_j^{\textbf{pqf}}$}{Vector of final predictive inputs to secondary controller of $j^{th}$ DER}

\section{Introduction}
\IEEEPARstart{P}{ower} electronic systems (PES) play a crucial role in enhancing efficiency, promoting sustainability and enabling flexibility. Achieving these objectives necessitates resilient control integrated with communication within PES, thus transforming PES into sophisticated cyber-physical system. The control framework of PES in this work, involves primary and secondary controllers. The conventional centralized secondary controllers (SCs) have limitations such as, high communication bandwidth, vulnerability to single-point failures and high computational complexity. To address these drawbacks, a highly reliable and scalable distributed secondary control (DSC) architecture is widely accepted, which only requires information from neighboring agents \cite{R1}. This complex network requires time-synchronized measurements. Global navigation satellite signals (GNSSs), such as GPS, GLONASS, BeiDou and Galileo, are the primary sources of time synchronization due to their worldwide coverage and high accuracy \cite{R2}. Intelligent electronic devices (IEDs) and merging units depend on GNSS for time transfer, using methods such as precision time protocol (PTP), inter-range instrumentation group time code B (IRIG-B), or one pulse per second (1PPS) \cite{R3}. However, integrating communication network exposes them to various constraints, like delays, data loss, and uncertain links \cite{R6}. These can cause delayed exchange of measurement/control signals among distributed energy resources (DERs), affecting system performance. 

The cyber-physical system further create opportunities for malicious attackers to launch coordinated cyber attacks. Among several cyber attacks \cite{R5,R5a}, this paper focuses on time delay-based cyber attacks, which can be strategically introduced into the control system by an adversary \cite{R7}. The time-synchronization attacks (TSAs) are a new kind of attack, which can manipulate the timing signals by corrupting the GNSS signals. Attackers can use a receiver-spoofer mechanism \cite{R9}, where the spoofer itself is a GPS receiver. Both space-based time synchronization (SBTS) and network-based time synchronization (NBTS) mechanisms \cite{R10} lack integrated security controls and have been accounted as highly vulnerable to TSAs \cite{R12}. This leads to false measurements and inaccurate time stamps, severely affecting the stability of the system.

The massive importance of time synchronized real-time measurements in cyber-physical networks makes it a valuable target to adversaries. Moreover, since PES have low system inertia and high response speed, the impact of these attacks are more significant than in bulk power systems. Therefore, making it crucial to design controllers that can withstand such cyber attacks within real-time operational constraints. In prior works, such as \cite{R23} and \cite{R24}, optimization-based methods were proposed for enhancing microgrid dynamic performance under communication delays. Nevertheless, these techniques come with notable computational overhead, especially in complex networks, and can be sensitive to initial conditions, potentially yielding suboptimal results. Another approach, as seen in \cite{R25}, employs predictive control theory, demanding a substantial amount of modeling knowledge. The requirement of observer/estimator in this scheme, increases the complexity further. Moreover, these schemes often struggle to establish resilience to unknown dynamics, risking performance degradation or instability. Furthermore, \cite{R26} introduces an anomaly-based scheme to detect the presence of TSAs and other attacks. However, this scheme necessitates a training phase, potentially entailing high memory and critical data requirements. Data-driven methods like these may require hyperparameter tuning and might encounter overfitting issues. While TSA detection schemes have been investigated in \cite{R27} and \cite{R28}, they lack a mitigation strategy to ensure stable PES operation during delays. Therefore, the existence of numerous distinct strategies to individually address data availability attacks, which often entail complex modeling or training approaches, motivated our proposal of a unified approach, capable of effectively mitigating all forms of such attacks. For this, we exploit the science of semantics to decipher a novel delay-aware semantic sampling scheme in this paper. Semantic principles have gained traction in various domains, including communication systems \cite{ref1} and networked intelligent systems \cite{ref2}. In speech recognition, semantics improves accuracy and efficiency in transforming spoken language into text \cite{ref3}. The post-5G era sees semantics shaping the future of wireless networks \cite{ref4}. For comprehensive insights into semantic communication and its applications, interested readers are encouraged to refer \cite{ref5}.

Real-time systems, such as smart grids and networked systems, rely on an automated sense-compute-actuate cycle for decision-making. The effectiveness of the connectivity in these systems hinges on the provision of \textit{right information to the right place at the right time}. During data availability attacks, our proposed delay-aware semantic sampling scheme addresses the challenge of real-time control operation and stability due to missing samples by employing \textit{semantic communication \& sampling}. The proposed scheme furnishes delay-compensation signals to the controller locally by rectifying the above mentioned missing samples through a semantic reconstruction process. This approach harnesses semantic attributes, namely value, freshness, and relevance, which are governed by factors like prioritization of the most significant signal for estimation, age of information (AoI), and reconstruction error, respectively. These semantics attributes tune the reconstruction process by extraction of \textit{significant} information from the dynamics of inner control loops through semantic sampling. These reconstructed signals are subsequently provided at a local level to SCs, effectively mitigating delays introduced by adversaries through data availability attacks. This distributed learning approach enhances the reliability and timeliness of information flow within real-time systems, enhancing the overall performance and resilience of PES.

In particular, the main contributions and benefits of this work are highlighted as:

\begin{itemize}
    \item The proposed delay-aware semantic sampling scheme, exploits significant information extracted from the inner control loop dynamics to provide reconstructed signals to local SC, facilitating delay compensation. This strategic approach minimizes redundant data transmissions. 
    \item The proposed scheme in this work, is robust against latency attacks, data dropouts and TSAs. It also guarantees the SC objectives are met under such attack scenarios. 
    \item The proposed delay-aware semantic sampling scheme embraces distributed approach, in contrast to complex centralized methods requiring intricate coordination between numerous components. Here, individual DERs independently handle local delay compensation, streamlining operations and enhancing manageability.
    \item The proposed scheme in this work, is model-agnostic. This simplifies implementation by eliminating the need for numerous device-specific models.
    \item Unlike training-based approaches that demand substantial computational resources, extensive datasets, and meticulous hyperparameter tuning, our approach operates without the need for training. It also does not have any additional hardware requirements.
\end{itemize}

The remainder of this paper is organized as: the science and relevance of semantics is explained in Section II. A brief description on modeling of cyber-physical PES is provided in Section III. The description, challenges and modeling of data availability attacks are illustrated in Section IV. The novel delay-aware semantic sampling approach is presented in Section V. The real-time simulation testbed setup and the performance evaluation of the proposed delay-aware semantic sampling scheme, is presented in Section VI. Finally, Section VII encapsulates the concluding remarks and future work.

\section{Science and Relevance of Semantics}
The term ``\textit{semantics}" originated from the ancient Greek word ``\textit{semantikos}", meaning \textit{significant}, and has evolved to refer to ``\textit{meaning}" in the context of languages. However, in this work, the term ``\textit{semantics}" is used in its original sense of ``significance" with regards to information. This approach recognizes that the relevance of information can vary depending on the application.  
\begin{figure}[h!]
        \centering
	    \includegraphics[clip, trim=0.5cm 2.8cm 3.8cm 0.5cm, width=1\linewidth]{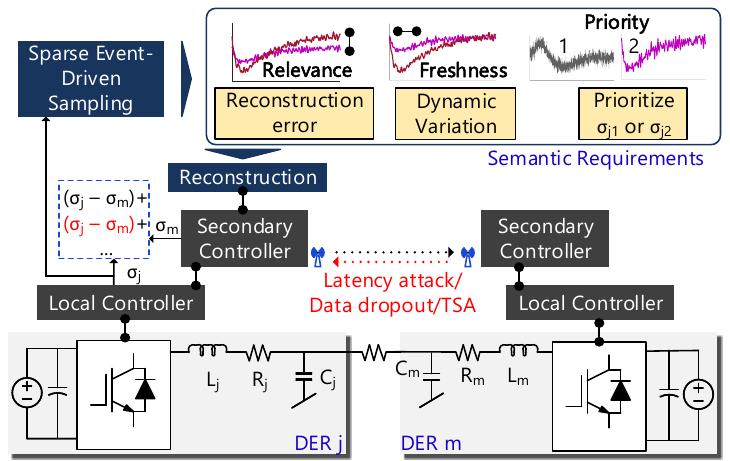}
	    \caption{Semantic information exchange and estimation in PES -- sparse event-driven sampling from local error measurements   steer the estimation and reconstruction process during latency attack/data dropout/TSAs.}
	    \label{fig:All}
\end{figure}

In semantic sampling, the three attributes of evaluating the criticality/significance of information are freshness, value, and relevance. Their definitions are as follows:
\begin{itemize}
\item \texttt{Freshness} refers to sending new updates at \textit{the right time}. It is defined as the time for the newest sample of information to reach from the source to the destination. Considering $\mathrm{u(t)}$ to be the timestamp of the latest packet received at destination by time $\mathrm{t}$, freshness is expressed as $\mathrm{F}(\mathrm{t})=\mathrm{t}-\mathrm{u(t)}$. \item \texttt{Value} refers to providing timely and \textit{right piece of information} to the \textit{right point of computation}, particularly in cyber-physical and hierarchical control systems. It defines \texttt{Priority} of information. \item \texttt{Relevance} involves generating the \textit{right piece of information} by sampling. It measures the extent of change in a process since the last recorded sample.
\end{itemize}

Based on the semantic requirements described above, we exploit it in the sampling and reconstruction process of new signals for each DER locally in PES, as shown in Fig. \ref{fig:All}.
As a result, the key focus is on steering the accuracy of estimation amid latency attacks, data dropouts and TSAs. Additionally, the semantic attributes i.e, relevance, freshness and priority are governed by reconstruction error, dynamic variation and prioritization of the most significant local signal to be used for estimation, respectively. Therefore, the semantic models pave way towards a standardized mechanism to represent and interpret from the relevant data collected from various devices and sensors across the network.

%lbrt
\begin{figure*}[t!]
        \centering
	     \includegraphics[clip, trim=0.5cm 3cm 6.2cm 0.5cm, width=0.9\textwidth]{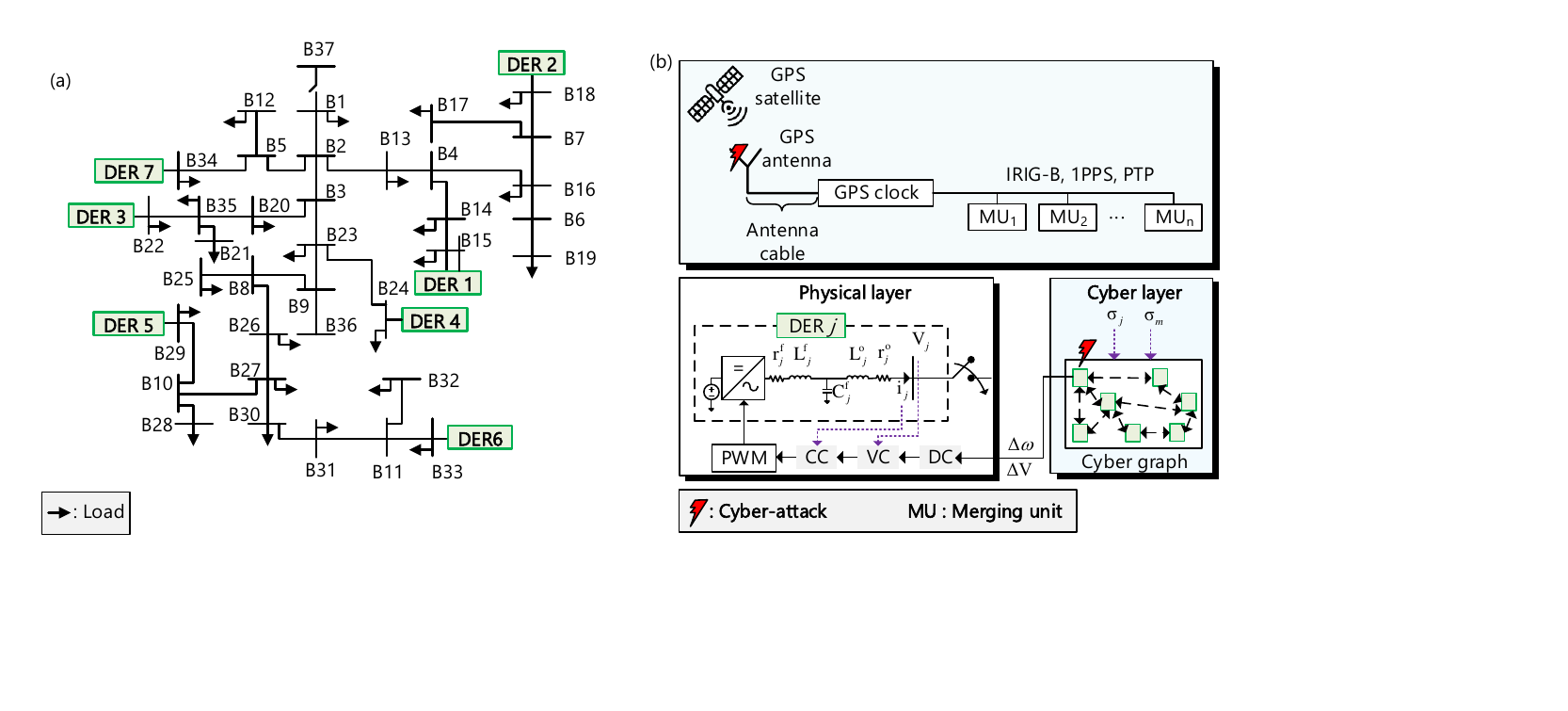}
	    \caption{(a) The modified IEEE 37-bus islanded AC distribution system powered by seven DERs is shown. (b) The block diagram of cyber-physical DER with primary and DSC architecture is presented. The DSC receives local measurements ($\mathrm{\sigma}_j$) and neighbouring measurements ($\mathrm{\sigma}_m$) as input to generate frequency and voltage correction terms ($\Delta \omega$ and $\Delta \mathrm{V}$). Note that the merging units (MUs) receive the timing information from GPS satellite. These time-stamped measurements are then used by the controllers for generating control signals, which can directly affect the control operation of the system.}
	    \label{fig:DER}
\end{figure*}

\section{Modeling Preliminaries}
\subsection{Physical Framework}
To demonstrate the modeling and control framework of a PES, the modified IEEE 37-bus system is presented in Fig. \ref{fig:DER}(a), with distributed loads powered by seven DERs. In the considered system, each DER can be represented by a DC source (denoting an energy storage system), DC/AC converter, LC filter ($\mathrm{r^f}$, $\mathrm{L^f}$, $\mathrm{C^f}$) and RL output impedance ($\mathrm{r^o}$, $\mathrm{L^o}$). The $d-q$ axis control framework comprises of inner control loops (voltage control (VC) and current control (CC)), cascaded with the primary droop control (DC) loop, as shown in Fig. \ref{fig:DER}(b). The merging units transmit the time-synchronized measurements (facilitated by GPS) to these controllers for the controller operation. As shown in Fig. \ref{fig:DER}(b), the GPS clock offers synchronized measurements of time by IRIG-B, PTP or 1PPS. The adopted frequency and voltage droop are:
\vspace{-0.2cm}
\begin{equation}
    \label{EQ1}
    {{\omega }_{j}^*}(t)=\omega_{\mathrm{nom}}-\mathrm{m}^{\mathrm{p}}_{j}{\mathrm{P}_{j}}\mathrm{(t)}
\end{equation}
\begin{equation}
    \label{EQ2}
    \mathrm{V}^{\mathrm{d*}}_{j}\mathrm{(t)}=\mathrm{V_{nom}}-\mathrm{n}^{\mathrm{q}}_{j}{\mathrm{Q}_{j}}\mathrm{(t)}\hspace{0.2cm},\hspace{0.2cm} \mathrm{V}^{\mathrm{q*}}_{j}\mathrm{(t)}=0
\end{equation}
where, the subscript `$j$' represents the parameters associated to $j^{th}$ DER. The terms $\omega_{\mathrm{nom}}$ and $\mathrm{V}_{\mathrm{nom}}$ are the nominal frequency and voltage of the AC system, respectively. The local reference frequency and voltage of a DER are ${{\omega }_{j}^*}$ and $\mathrm{V}^{\mathrm{dq*}}_{j}$. Here, $\mathrm{V}^{\mathrm{dq*}}_{j}\mathrm{(t)}=[\mathrm{V}^{\mathrm{d*}}_{j}\mathrm{(t)} \hspace{0.2cm} \mathrm{V}^{\mathrm{q*}}_{j}\mathrm{(t)}]^{\mathrm{T}}$. The active and reactive power droop coefficient are  $\mathrm{m}^{\mathrm{p}}$ and $\mathrm{n}^{\mathrm{q}}$, respectively. More information about its control layer modeling can be referred from \cite{R30}. Since primary control inherently results in non-zero steady-state error, the DSC scheme is integrated, as in Fig. \ref{fig:DER}(b), described in the next subsection.

\subsection{Cyber Framework}
Let us consider PES with $M$ power electronic-interfaced DERs in a sparsely-connected DSC based communication network. These DERs are termed as agents/nodes in cyber layer and are represented as $\textbf{x} = \{\mathrm{x}_1, \mathrm{x}_2, … , \mathrm{x}_M\}$. These agents are linked to their neighbouring agents by edges $\mathbf{E}$ via an associated adjacency matrix, ${\mathbf A_\text{G}} = [\mathrm{a}_{jm}]\in{\textbf{R}^{N\times{N}}}$. The neighbours to $j^{th}$ agent is represented as, $\mathrm{N}_{j} = \{ {m} \ | \ (\mathrm{x}_m, \mathrm{x}_j) \in \mathbf{E} \} $. Here, the communication weight $\mathrm{a}_{jm}$ (from agent $m$ to agent $j$) is modeled as: $\mathrm{a}_{jm} >$ 0, if ($\mathrm{x}_j$, $\mathrm{x}_m$) $\in$ $\mathbf{E}$. If there is no cyber link between $\mathrm{x}_j$ and $\mathrm{x}_m$, then $\mathrm{a}_{jm}$ = 0. Any agent sends/receives the information from the neighbouring agent(s) i.e, $\mathrm{\sigma}_m=[\mathrm{\omega}_m \hspace{0.2cm} \mathrm{m}_m^{\mathrm{p}}\mathrm{P} \hspace{0.2cm} \mathrm{n}_m^{\mathrm{q}}\mathrm{Q}]^\mathrm{T}$. The matrix representing incoming information can be given as, $\mathbf{D}_\text{in} = \texttt{diag}\{\mathrm{d}_j^{\mathrm{in}}\}$, where $\mathrm{d}_j^{\mathrm{in}}=\sum_{m\in \mathrm{N}_{j}}\mathrm{a}_{jm}$. Combining the sending and receiving end information into a single matrix, we obtain Laplacian matrix  $\mathbf{L}$ = [$\mathrm{l}_{jm}$], where $\mathrm{l}_{jm}$ are its elements defined such that, $\mathbf{L}$ = ${\mathbf{D}_\text{in}} – {\mathbf{A}_\text{G}}$. According to \cite{R30}, local reference frequency and voltage of DER, as expressed in \eqref{EQ1} and \eqref{EQ2}, are re-defined as:
\vspace{-0.2cm}
\begin{equation}
    \label{EQ3}
    {{\omega }_{j}^*}\mathrm{(t)}=\omega_{\mathrm{nom}}-\mathrm{m}^{\mathrm{p}}_{j}{\mathrm{P}_{j}}\mathrm{(t)}+\Delta{\omega \mathrm{c}}_{j}\mathrm{(t)}
\end{equation}
\begin{equation}
    \label{EQ4}   \mathrm{V}_{j}^{\mathrm{d*}}\mathrm{(t)}=\mathrm{V}_{\mathrm{nom}}-\mathrm{n}^{\mathrm{q}}_{j}{\mathrm{Q}_{j}}\mathrm{(t)}+\Delta \mathrm{Vc}_{j}\mathrm{(t)}\hspace{0.2cm}
\end{equation}
where, $\Delta {\mathrm{\omega}\mathrm{c}}$ and $\Delta {\mathrm{Vc}}$ are the frequency and voltage correction terms from the SC, expressed as: 
\vspace{-0.1cm}
\begin{equation}
\label{EQ5}
\begin{aligned}
    \Delta{\omega \mathrm{c}}_{j}\mathrm{(t)}=&-\mathrm{H}_1(s)[\mathrm{\omega_{nom}}-\omega_j\mathrm{(t)}+\\ &\mathrm{g}_j\sum\limits_{m\in {\mathrm{N}_{j}}}{{\mathrm{a}_{jm}}\left( \mathrm{\omega}_m\mathrm{(t)}-\mathrm{\omega}_j\mathrm{(t)} \right)}+\\ &\mathrm{g}_j\sum\limits_{m\in {\mathrm{N}_{j}}}{{\mathrm{a}_{jm}}\left(\mathrm{m}_m^\mathrm{p}\mathrm{P}_m\mathrm{(t)}-\mathrm{m}_j^\mathrm{p}\mathrm{P}_j\mathrm{(t)} \right)}]
\end{aligned}
\end{equation}
\vspace{-0.1cm}
Similarly,
\begin{equation}
\label{EQ7}
    \Delta \mathrm{Vc}_{j}\mathrm{(t)}=-\mathrm{H_2(s)}[\mathrm{g}_j\sum\limits_{m\in {\mathrm{N}_{j}}}{{\mathrm{a}_{jm}}\left(\mathrm{n}_m^\mathrm{q}\mathrm{Q}_m\mathrm{(t)}-\mathrm{n}_j^\mathrm{q}\mathrm{Q}_j\mathrm{(t)} \right)}]
\end{equation}
where, $\mathrm{H_1(s)}$ and $\mathrm{H_2(s)}$ are PI controllers for frequency restoration along with proportional active power sharing; and proportional reactive power sharing, respectively. The local control input of SC can be given by:
\vspace{-0.1cm}
\begin{equation}
    \textbf{u}_j\mathrm{(t)}=\mathrm{g}_j\sum\limits_{m\in {\mathrm{N}_{j}}}\underbrace{{{\mathrm{a}_{jm}}\left( \boldsymbol{\sigma}_m\mathrm{(t)}-\boldsymbol{\sigma}_j\mathrm{(t)} \right)}}_{\textbf{e}_{jm}\mathrm{(t)}}
\end{equation}
where, $\textbf{u}_j=[\mathrm{u}_j^\mathrm{p} \hspace{0.2cm} \mathrm{u}_j^\mathrm{q}]^{\mathrm{T}}$, $\textbf{e}_{jm}=[\mathrm{e}_{jm}^\mathrm{p} \hspace{0.2cm} \mathrm{e}_{jm}^\mathrm{q}]^{\mathrm{T}}$, depending on the elements in $\boldsymbol{\sigma}$; and $\mathrm{g}_j$ is the convergence parameter. 

These information exchanges can be limited by data availability cyber-attacks, which then aggravates the system monitoring and controllability due to missing information, as explained in the next section.

\section{Overview of Data Availability Attacks}

\subsection{Latency Attacks and Data Dropouts}
\textbf{Description and challenges:}
Communication time-delays are an inherent part of any communication system encompassing four primary components: propagation delay, transmission delay, processing delay, and queuing delay \cite{R31}. In the DSC architecture, real-time periodic communication is essential for efficient operation. However, data congestion can introduce unpredictable delays, influenced by factors like cyber sampling rate, data volume, and cyber graph connection. These delays, ranging from milliseconds to seconds, can disrupt system operation if they exceed SC operational time limits \cite{R32}. Preventive measures are crucial to avoid missed updates that could lead to oscillatory instability or system failure.

\begin{figure}[h!]
        \centering
	    \includegraphics[clip, trim=0.7cm 2.3cm 3.9cm 0.7cm, width=1\linewidth]{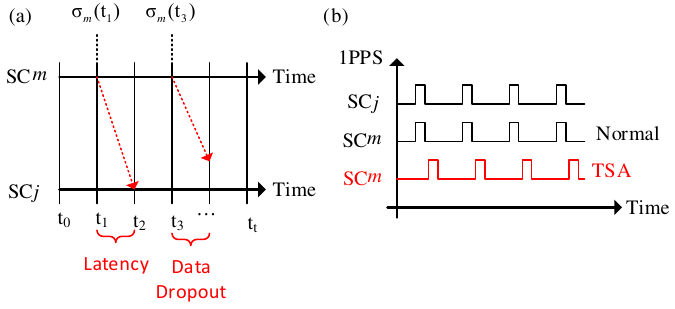}
	    \caption{\textbf{(a)} Latency attack and data dropout; and \textbf{(b)} TSA.}
	    \label{fig:Attacks}
\end{figure}

Furthermore, cyber attackers can exacerbate issues by intentionally adding time delays to critical messages, known as latency attacks (as shown in Fig. \ref{fig:Attacks}(a)). This can severely impact time-critical information transfer between SCs. Network congestion can also cause frequent data dropouts (as shown in Fig. \ref{fig:Attacks}(a)), further compromising dynamic performance.\\
\textbf{Attack model:}
The DSC fundamentally relies on the accurate transmission of data from neighboring agents. Latency attacks, which introduce falsifications in timing signals, pose a substantial threat to the operational stability of the system. These attacks can exert a profound influence on the control laws that govern the behavior of cyber-physical PES, potentially leading to significant deviations from desired performance.

In this context, considering the neighbors of the $j^{th}$ agent be denoted as $\mathrm{N}_{j} = { {m} \ | \ (\mathrm{x}_m, \mathrm{x}_j) \in \mathbf{E} } $. The local control input of the SC, when subjected to latency attack is:
\vspace{-0.1cm}
\begin{equation}
\label{EQ8}
    \textbf{u}_j^\mathrm{L}(t)=\mathrm{g}_j\sum\limits_{m\in {\mathrm{N}_{j}}}{{\mathrm{a}_{jm}}\left( \boldsymbol{\sigma}_m(\mathrm{t}-\mathrm{\tau}_m)-\boldsymbol{\sigma}_j(\mathrm{t}-\mathrm{\tau}_j) \right)}
\end{equation}
where, $\mathrm{\tau}_j$ and $\mathrm{\tau}_m$ are the delays from the local and neighbouring agents. By Leibnitz formula, the delayed parameter can be expressed as, $\mathrm{\sigma(t-\tau)=\sigma(t)-\int_{t-\tau}^{t}\dot{\sigma}(s)ds}$. For a delay of $\mathrm{\tau}_m$, substituting this in \eqref{EQ8}, we obtain, $\mathrm{\dot{\sigma}(t)=-\textbf{L}\sigma(t)-\textbf{A}\int_{t-\tau_m}^{t}\dot{\sigma}(s)ds}$. Similarly, the expression for local delay can also be obtained. For a fixed, undirected and connected cyber graph, the equilibrium is reached, if and only if, $0<\mathrm{\tau}<\mathrm{\frac{\pi}{2\lambda_{max}\mathbf{L}}}$, with $\lambda_{max}$ being the largest eigenvalue of $\mathbf{L}$. Thus, the communication delay ($\mathrm{\tau}$) must be bounded inside these limits to obtain $\mathrm{\dot{\sigma}(t)=0}$. \begin{figure}[b]
        \centering
	    \includegraphics[clip, trim=0.5cm 1.3cm 3.2cm 0.7cm, width=0.8\linewidth]{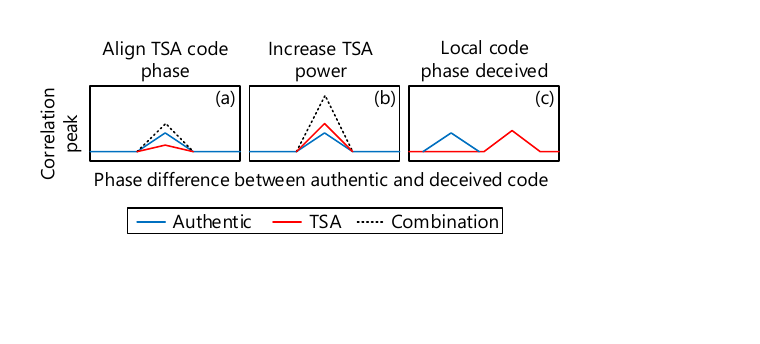}
	    \caption{Spoofing procedure for TSA. \textbf{(a)} Aligning TSA code phase with authentic one; \textbf{(b)} initiating attack by increasing TSA signal power; and \textbf{(c)} gradual alteration of the victim's code to introduce timing error.}
	    \label{fig:TSA_P}
\end{figure} \begin{figure*}[b!]
        \centering
	    \includegraphics[clip, trim=0.7cm 3.3cm 8cm 0.5cm, width=1\linewidth]{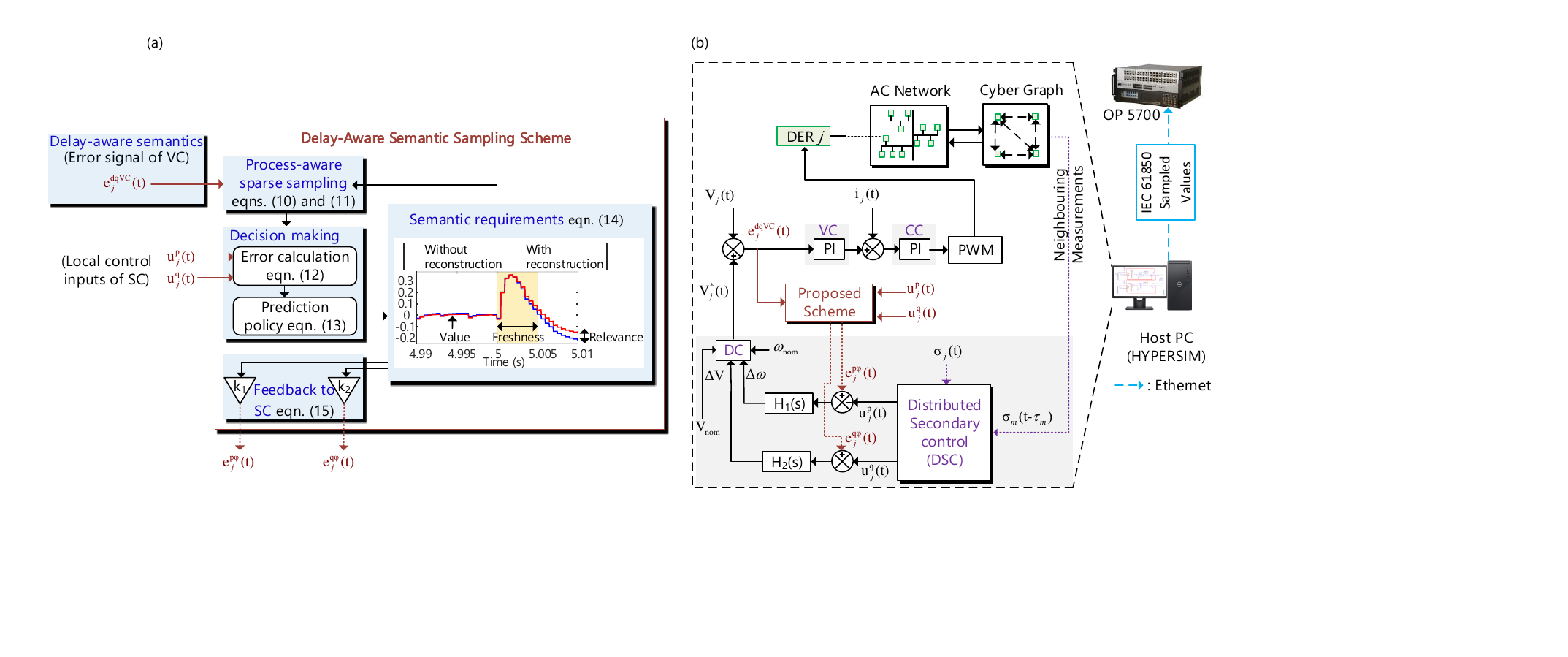}
	    \caption{\textbf{(a)} Proposed delay-aware semantic sampling scheme. \textbf{(b)} Deployment of the proposed scheme in real-time simulation testbed. The testbed is interfaced with Ethernet to facilitate establishment of IEC 61850 sampled values protocol.}
	    \label{fig:Proposed}
    \end{figure*} 
    
\subsection{Time Synchronization Attacks (TSAs)}
Recently, there has been a significant upsurge in TSAs, which is becoming a growing concern across various sectors. This concern arises from the susceptibility of GPS signals to compromise by unintentional sources like radio frequency (RF) interference and space weather events such as solar flares. Such interference can result in timing errors or even complete signal loss, posing critical risks to time-sensitive applications.

Beyond unintentional disruptions, GPS receivers in devices like substation clocks or merging units face vulnerability to deliberate attacks by malicious actors. For instance, GNSS signals, transmitted by satellite constellations in medium earth orbit (MEO), exhibit low power levels, with a power density of fW/$\mathrm{m}^2$ ($10^{-15}$ W/$\mathrm{m}^2$) upon reaching the Earth's surface \cite{R35}. To illustrate, this is akin to observing a 25 W light bulb from a distance of 10,000 miles. Consequently, these signals become susceptible to blocking or jamming over extensive areas through low-power terrestrial transmitters, effectively saturating the GNSS signal spectrum with noise or an unmodulated carrier.

While the \textit{blocking/jamming attack} is relatively straightforward to detect due to a complete time loss, \textit{spoofing} of GNSS signals presents a more challenging threat. Spoofing entails the broadcast of fraudulent GPS signals or the rebroadcasting of GPS signals captured at a different time-step at the target receiver (as shown in Fig. \ref{fig:Attacks}(b)). This deceptive manipulation can lead to time synchronization loss, diminishing network synchronization performance and, consequently, reducing the stability and reliability of the PES.

\textbf{Attack model:} 
GNSS timing relies on phase of pseudo-random noise (PRN) codes within received signals \cite{R27}. To manipulate timing results, TSA signals must alter these values, as shown in Fig. \ref{fig:TSA_P}. Initially, the attacker aligns TSA signal code phases with authentic ones, maintaining a relatively low signal power, as shown in Fig. \ref{fig:TSA_P}(a). Once alignment is achieved, the attack can be initiated at any time by increasing TSA signal power while slowly shifting code phases away, as shown in Fig. \ref{fig:TSA_P}(b). Tracking loops will then lock onto TSA correlation peaks due to their higher power, enabling TSA signals to dominate all tracking loops without causing them to lose lock on signals. Simultaneously, the victim's code undergo gradual alteration, introducing errors into the timing results, as shown in Fig. \ref{fig:TSA_P}(c). More details regarding TSA modelling can be referred from \cite{R12}.

In such events the time-stamped data of the victim node, $\boldsymbol{\sigma}_m\mathrm{(t)}$ is manipulated by an offset of $\mathrm{nT_s}$ samples, the resultant attacked information can be expressed as:
\vspace{-0.1cm}
\begin{equation}
    \sigma_m^\mathrm{T}\mathrm{(t)}=\sigma_m\mathrm{(t \pm nT_s)}
\end{equation}
Whether the adversary chooses to add or subtract these $\mathrm{nT_s}$ samples, the timing information is compromised, which can lead to time synchronization loss. Consequently, inaccurate time stamps which reverberate through the entire system, exerts a detrimental influence on the precise coordinated operation of DERs within the PES. Within the SC, the integrator accumulates error based on the latest available data. The gradual accumulation of error over time can be substantial which can steer the control system away from its intended setpoint. The control system may exhibit undesirable behaviors such as oscillations, overshooting, etc. As the error accumulates and amplifies, it has the potential to induce instability.

In PES, the above-mentioned cyber attacks can result in a host of problems ranging from sub-optimal operating conditions to outright instability of the system. This instability may even cause inadvertent disconnection of sources/ loads, leading to partial/full shutdown of the system, thereby jeopardizing the security of electrical supply. To address these challenges, it is crucial to implement a robust control system to handle unpredictable delays. Therefore, efforts are accumulated to work in this direction, presented in the next section.

\section{Proposed Delay-Aware Semantic Sampling}
As previously mentioned, the term \textit{semantics} refers to the significance of information. By incorporating the concept of information semantics, this paper aims to provide a more nuanced and comprehensive understanding of the role of information in decision-making during delays in networked PES. The contextual representation of semantics in PES refers to capturing the attributes of inner control loop signals, such as timeliness and value, to reconstruct significant information necessary for delay compensation in scenarios involving random delay attacks. In distributed control of AC distribution systems, timely consensus negotiation among agents is crucial for global frequency regulation and proportional active/reactive power sharing. Semantic-aware transmission, which respects the time-dependent value of signals, is essential to ensure achieving the SC objectives. \RestyleAlgo{ruled}
\SetKwComment{Comment}{/* }{ */}

\begin{algorithm}[t!]
\caption{Proposed delay-aware semantic sampling scheme at $j^{th}$ DER}\label{alg:two}
\textbf{Inputs:}  Error signals provided to VC ($\textbf{e}_j^{\textbf{dqVC}}\mathrm{(t)}$), length of window (W), downsampling factor (D), local control inputs to SC ($\textbf{u}_j^{\textbf{pq}}\mathrm{(t)}$), tunable parameter ($\alpha$), controller time constant of $\mathrm{H_1(s)}$ and $\mathrm{H_2(s)}$ PI control loops ($\mathrm{T=K_p/K_i}$), triggering moment ($\mathrm{t}_a$), tunable gains ($\mathrm{k_1}$ and $\mathrm{k_2}$) \\

\textbf{Signals:} impulse response ($\delta[w]$), downsampled signal ($\textbf{e}_j^{\textbf{dqD}}\mathrm{(t)}$), error fed to prediction policy ($\textbf{e}_j(\mathrm{t}_a)$), reconstructed signals ($\textbf{e}_j^{\textbf{R}}(\mathrm{t}_a)$), final predictive inputs to SC ($\textbf{u}_j^{\textbf{pqf}}\mathrm{(t)}$), freshness ($\mathrm{F}\mathrm{(t)}$), relevance ($\mathrm{R(t)}$), $\mathrm{u(t)}$ is the timestamp of the latest packet received at destination by time $\mathrm{t}$. \\

\textbf{Note:} $\textbf{e}_j^{\textbf{dqVC}}\mathrm{(t)}=[\mathrm{e}_j^{\mathrm{dVC}}\mathrm{(t)} \hspace{0.2cm} \mathrm{e}_j^{\mathrm{qVC}}\mathrm{(t)}]^{\mathrm{T}}$, $\textbf{e}_j^{\textbf{dqD}}\mathrm{(t)}=[\mathrm{e}_j^\mathrm{dD}(t) \hspace{0.2cm} \mathrm{e}_j^\mathrm{qD}(t)]^{\mathrm{T}}$,
$\textbf{u}_j\mathrm{(t)}=[\mathrm{u}_j^\mathrm{p}\mathrm{(t)} \hspace{0.2cm} \mathrm{u}_j^\mathrm{q}\mathrm{(t)}]^{\mathrm{T}}$,
$\textbf{e}_j(\mathrm{t}_a)=[{\mathrm{e}_j^\mathrm{p}(\mathrm{t}_a) \hspace{0.2cm} \mathrm{e}_j^\mathrm{q}(\mathrm{t}_a)}]^{\mathrm{T}}$,
$\textbf{e}_j^{\textbf{R}}(\mathrm{t}_a)=[\mathrm{e}_j^{\mathrm{Rp}}(\mathrm{t}_a) \hspace{0.2cm} \mathrm{e}_j^{\mathrm{Rq}}(\mathrm{t}_a)]^{\mathrm{T}}$,
$\textbf{u}_j^{\textbf{pqf}}\mathrm{(t)}=[\mathrm{u}_j^{\mathrm{pf}}\mathrm{(t)} \hspace{0.2cm} \mathrm{u}_j^{\mathrm{qf}}\mathrm{(t)}]^{\mathrm{T}}$ \\

\KwData{$i \geq 0$}
$I \gets i$\;
// Initialize: $\mathrm{F}(\mathrm{t}) = 0$, $\mathrm{R(t)} \neq 0$ \\ 
\While{$I \neq 0$}{
\vspace{0.1cm}
\eIf{($\mathrm{F}(\mathrm{t}) = 0 \hspace{0.2cm} \&\& \hspace{0.2cm} \mathrm{R(t)} \neq 0$)}{
    \vspace{0.1cm}
    // Compute freshness \eqref{EQ14}: $\mathrm{F}(\mathrm{t})=\mathrm{t}-\mathrm{u(t)}$ \\
    \vspace{0.1cm}
    // Process-aware sparse sampling \eqref{EQ10}, \eqref{EQ11}
    $\mathrm{e}_j^\mathrm{dD}=\sum_{w=0}^{\mathrm{W-1}}\mathrm{e}_j^{\mathrm{dVC}}[\mathrm{nD}-w].\delta[w]$\\
    $\mathrm{e}_j^\mathrm{qD}=\sum_{w=0}^{\mathrm{W-1}}\mathrm{e}_j^{\mathrm{qVC}}[\mathrm{nD}-w].\delta[w]$ \\
    \vspace{0.1cm}
    // Error generated for prediction policy \eqref{EQ12} \\
    $\textbf{e}_j(\mathrm{t}_a)= [{\mathrm{e}_j^\mathrm{dD}(\mathrm{t}_a) \hspace{0.1cm} \mathrm{e}_j^\mathrm{qD}(\mathrm{t}_a)}]-\textbf{u}_j$ \\
    \vspace{0.1cm}
    // Triggers generation with the prediction policy condition \eqref{EQ13} \\
  \eIf{($||\textbf{e}_j(\mathrm{t}_a)||>\alpha||e^{-\mathrm{t/T}}.[\mathrm{e}_j^{\mathrm{dVC}} \hspace{0.1cm} \mathrm{e}_j^{\mathrm{qVC}}]||$)}{
    \vspace{0.1cm}
    // Reconstruction of signals \\
    $\textbf{e}_j^{\textbf{R}}(r\mathrm{t}_a+\Gamma)$=$\textbf{e}_j(r\mathrm{t}_a)$; 0$\leq \Gamma < \mathrm{t}_a$ and r=0,1,2,...\\
    \vspace{0.1cm}
    // Reconstructed signals fed back to SC with tunable gains \eqref{EQ15} \\
    $\mathrm{e}_{j}^{\mathrm{p}\varphi}(\mathrm{t}_a)=\mathrm{k_1}\mathrm{e}_j^{\mathrm{Rp}}(\mathrm{t}_a) \hspace{0.1cm}$\\
    $\mathrm{e}_{j}^{\mathrm{q}\varphi}(\mathrm{t}_a)=\mathrm{k_2}\mathrm{e}_j^{\mathrm{Rq}}(\mathrm{t}_a)$\\
    
    \vspace{0.1cm}
    // Final predictive inputs fed to the SC for delay compensation \eqref{EQ16}\\ $\mathrm{u}_j^{\mathrm{pf}}\mathrm{(t)}=\mathrm{u}_j^\mathrm{p}\mathrm{(t)}+\mathrm{e}_{j}^{\mathrm{p}\varphi}(\mathrm{t}_a) \hspace{0.1cm}$\\
    $\mathrm{u}_j^{\mathrm{qf}}\mathrm{(t)}=\mathrm{u}_j^\mathrm{q}\mathrm{(t)}+\mathrm{e}_{j}^{\mathrm{q}\varphi}(\mathrm{t}_a)$ \\
    \vspace{0.1cm}
    // Compute relevance \eqref{EQ14}: $\mathrm{R(t)}=\textbf{e}_j-\textbf{e}_j^{\textbf{R}}$ \\
  }{// No reconstruction: $\mathrm{R(t)=0}$ \\
  }
  }{// Compute freshness \eqref{EQ14}: $\mathrm{F}(\mathrm{t})=\mathrm{t}-\mathrm{u(t)}$ \\
  }
}
\end{algorithm} 

Time-critical applications like smart grids and networked control require a restructured message transfer system due to the huge amount of data involved. Hence, this paper proposes a semantic sampling architecture as shown in Fig. \ref{fig:Proposed}, that generates and transmits the right amount of data to the right place at the right time. This includes following steps:

\renewcommand{\theenumi}{\roman{enumi}}%
\begin{enumerate}
  \item \textit{Delay-aware semantics:} To comprehend the proposed approach, it is crucial to apply the PI consensusability law \cite{R38} to anticipate the physical layer semantics using the response of each control loop under disturbances. This proposed scheme is local to each SC and firstly extracts significant information from the error signal corresponding to the VC ($\textbf{e}_j^{\textbf{dqVC}}\mathrm{(t)}$). Here, $\textbf{e}_j^{\textbf{dqVC}}\mathrm{(t)}=[\mathrm{e}_j^{\mathrm{dVC}}\mathrm{(t)} \hspace{0.2cm} \mathrm{e}_j^{\mathrm{qVC}}\mathrm{(t)}]^{\mathrm{T}}$.\\
  
  \item \textit{Process-aware sparse sampling \cite{new}, \cite{new2}:} The signal $\textbf{e}_j^{\textbf{dqVC}}\mathrm{(t)}$, is then downsampled (shown in Fig. \ref{fig:Proposed}(a)), as:
  \vspace{-0.1cm}
\begin{equation}
\label{EQ10}
    \mathrm{e}_j^\mathrm{dD}=\sum_{w=0}^{\mathrm{W-1}}\mathrm{e}_j^{\mathrm{dVC}}[\mathrm{nD}-w].\delta[w]
\end{equation}
\vspace{-0.1cm}
\begin{equation}
\label{EQ11}
    \mathrm{e}_j^\mathrm{qD}=\sum_{w=0}^{\mathrm{W-1}}\mathrm{e}_j^{\mathrm{qVC}}[\mathrm{nD}-w].\delta[w]
\end{equation}
where, $\delta[w]$ is an impulse response, W is the length of window, D is the downsampling factor. Downsampling is a resampling technique that decreases the resolution of the incoming signal, typically used to minimize memory usage. However, in this study, it is performed to align the dynamic performance of error quantities fed to VC (i.e, $\mathrm{e}_j^{\mathrm{dVC}}\mathrm{(t)}$ and $\mathrm{e}_j^{\mathrm{qVC}}\mathrm{(t)}$) and error fed to SC (i.e, $\textbf{u}_j\mathrm{(t)}$). This crucial step aids in the synchronization of the multi-time scale error signals. This approach significantly lowers device energy consumption. This effect is rooted in the definition of energy consumption, which is the product of power consumption and processing time for each sample \cite{ref6}. Downsampling, by reducing the number of samples based on the D, decreases energy consumption as D increases. This is crucial particularly for low-power/energy-harvesting sensors, while also enabling efficient bandwidth utilization.\\

  \item \textit{Effective decision making:} The generated downsampled signals ($\mathrm{e}_j^\mathrm{dD}\mathrm{(t)}$ and $\mathrm{e}_j^\mathrm{qD}\mathrm{(t)}$) are compared with the local control inputs from the SC (i.e, $\mathrm{u}_j^\mathrm{p}\mathrm{(t)}$ and $\mathrm{u}_j^\mathrm{q}\mathrm{(t)}$), as shown in Fig. \ref{fig:Proposed}(a). The semantic prediction policy subsequently rebuilds the signals used for delay compensation (i.e, $\textbf{e}_j(\mathrm{t}_a)=[{\mathrm{e}_j^\mathrm{p}(\mathrm{t}_a) \hspace{0.2cm} \mathrm{e}_j^\mathrm{q}(\mathrm{t}_a)}]^{\mathrm{T}}$) as:
\vspace{-0.1cm}
\begin{equation}
\label{EQ12}
    \textbf{e}_j(\mathrm{t}_a)= [{\mathrm{e}_j^\mathrm{dD}(\mathrm{t}_a) \hspace{0.2cm} \mathrm{e}_j^\mathrm{qD}(\mathrm{t}_a)}]-\textbf{u}_j
\end{equation}
%\vspace{-0.1cm}
Additionally, the error is fed into the prediction policy stage to generate a signal that compensates for significant delays. The prediction policy condition is expressed as:
\vspace{-0.1cm}
\begin{equation}
    \label{EQ13}
    ||\textbf{e}_j(\mathrm{t}_a)||>\alpha||e^{-\mathrm{t/T}}.[\mathrm{e}_j^{\mathrm{dVC}} \hspace{0.2cm} \mathrm{e}_j^{\mathrm{qVC}}]||
\end{equation}
where, $\alpha$ is a tunable parameter, $\mathrm{T=K_p/K_i}$ is the controller time constant of $\mathrm{H_1(s)}$ and $\mathrm{H_2(s)}$ PI control loops. If the condition expressed in \eqref{EQ13} is met, triggers are produced. These triggers are utilized to reconstruct $\textbf{e}_j^{\textbf{R}}(\mathrm{t}_a)$ using a sample-and-hold circuitry, with $\mathrm{t}_a$ being the triggering instant. This is followed by evaluation of semantic attributes i.e, freshness ($\mathrm{F}\mathrm{(t)}$), value and relevance ($\mathrm{R(t)}$) defined as:
\vspace{-0.1cm}
\begin{equation}
\label{EQ14}
    \mathrm{F}\mathrm{(t)}=\mathrm{t}-\mathrm{u(t)} \hspace{0.2cm}, \hspace{0.2cm} \mathrm{R(t)}=\textbf{e}_j(t)-\textbf{e}_j^{\textbf{R}}(t)
\end{equation}
where, $\mathrm{u(t)}$ is the timestamp of the latest packet received at destination by time $\mathrm{t}$.

\item \textit{Feedback generation:} The resulting reconstructed signals are subsequently fed back to SC, with their tunable gains, $\mathrm{k_1}$ and $\mathrm{k_2}$, represented as:
\vspace{-0.1cm}
\begin{equation}
\label{EQ15}
    \mathrm{e}_{j}^{\mathrm{p}\varphi}(\mathrm{t}_a)=\mathrm{k_1}\mathrm{e}_j^{\mathrm{Rp}}(\mathrm{t}_a) \hspace{0.2cm}, \hspace{0.2cm} \mathrm{e}_{j}^{\mathrm{q}\varphi}(\mathrm{t}_a)=\mathrm{k_2}\mathrm{e}_j^{\mathrm{Rq}}(\mathrm{t}_a)
\end{equation}
where, $\textbf{e}_j^{\textbf{R}}(\mathrm{t}_a)=[\mathrm{e}_j^{\mathrm{Rp}}(\mathrm{t}_a) \hspace{0.2cm} \mathrm{e}_j^{\mathrm{Rq}}(\mathrm{t}_a)]^{\mathrm{T}}$. Finally these inputs are added to the control inputs of SC as:
\vspace{-0.1cm}
\begin{equation}
   \label{EQ16}
   \mathrm{u}_j^{\mathrm{pf}}\mathrm{(t)}=\mathrm{u}_j^\mathrm{p}\mathrm{(t)}+\mathrm{e}_{j}^{\mathrm{p}\varphi}(\mathrm{t}_a) \hspace{0.2cm}, \hspace{0.2cm} \mathrm{u}_j^{\mathrm{qf}}\mathrm{(t)}=\mathrm{u}_j^\mathrm{q}\mathrm{(t)}+\mathrm{e}_{j}^{\mathrm{q}\varphi}(\mathrm{t}_a)
\end{equation}
where, $\mathrm{u}_j^{\mathrm{pf}}$ and $\mathrm{u}_j^{\mathrm{qf}}$ are the final predictive inputs to the SC to compensate the delays.
\end{enumerate}
The control objectives of the proposed delay-aware semantic sampling scheme, may be summarized as:
    \begin{enumerate}[label=(\roman*)]
        \item To address delayed communication signals resulting from latency attacks, data dropouts, or TSAs by incorporating \textit{semantic} principles into the sampling process for each DER. This integration enables the generation of reconstruction signals (fed back to local SC), based on the inner control layer dynamics. 
        \item To evaluate the reconstruction phase by filtering significant events caused during data availability attacks. Considering dynamic variation, prioritization of signals and computation of reconstruction error, reconstruction signals are tuned to generate delay compensation signals.
    \end{enumerate} 
Thus, the scheme targets optimal information gathering, dissemination, and decision-making policies in cooperative networks, achieving jointly optimal performance. The convergence analysis of the proposed scheme is discussed further.

\subsection{Convergence Analysis of the Proposed Scheme}
Let $\texttt{o}(\mathrm{t}_a)$ denotes the triggered samples of the respective signals when triggering condition is met during data availability attacks. The proposed delay-aware semantic sampling scheme's convergence analysis is theoretically discussed and validated. Let the reconstructed signals ($\textbf{e}_j^{\textbf{R}}(\mathrm{t}_a)$) produce the triggered voltage correction term ($\Delta {\mathrm{Vc}}(\mathrm{t}_a)$) and frequency correction term ($\Delta {\omega \mathrm{c}}(\mathrm{t}_a)$) from the SC. Taking into account the triggered sampled measurements as:
\vspace{-0.1cm}
    \begin{equation}
        \hat{\Upsilon}_j(k)=\Upsilon_j(\mathrm{t}_a)
    \end{equation}
where, $k\in [\mathrm{t}_a,\mathrm{t}_{a+1}]$ and $\Upsilon_j=\{\Delta Vc_j, \Delta \omega \mathrm{c}_j\}$. Let us define 
\vspace{-0.1cm}
    \begin{equation}
    \label{X1}
        y_j(\mathrm{t}_a)=\hat{\Upsilon}_j(\mathrm{t}_a)-\frac{1}{N_j}\sum_{m \in N_j}\Upsilon_j^T(t) \forall j \in M
    \end{equation}
Let $\mathrm{t}_a^j, \forall a=1,2,...$ represent the triggering instants in the $j^{th}$ agent. $M$ is the total number of DERs in a network and $N_j$ is the neighbouring agents to $j^{th}$ agent. Consequently, the sampled control input becomes a piece-wise constant function, where $\hat{u}_j(k)=u_j(\mathrm{t}_a^{N_j})$ for $k \in [\mathrm{t}_a^{N_j},\mathrm{t}_{a+1}^{N_j})$. Considering the initial condition $\Upsilon(0)$, the iteration within the proposed delay-aware semantic sampling scheme for the $j^{th}$ agent is:
\vspace{-0.1cm}
    \begin{equation}
        \label{X2}
        \Upsilon_j(k+1)=\Upsilon_j(k)+\beta_ju_j(k)
    \end{equation}
Here, $\beta_j$ represents the step length. Employing a Lyapunov candidate function, denoted as $V(\Upsilon(k))=f(\Upsilon(k))-f(\hat{\Upsilon}(k))$ for the system in \eqref{X2}, it is trivial to deduce from \eqref{X1} to \eqref{X2} that $\Delta V(\Upsilon)=\Delta f(\Upsilon)$. For all $k \geq 0$,
\vspace{-0.1cm}
    \begin{equation}
    \label{X3}
        \Delta V \leq \sum_{j=1}^{M}\left \{ \beta_ju_j \left [ \sum_{m \in N_i}(\Upsilon_m-\hat{\Upsilon}_j)-u_j \right ] +\frac{M}{2}\beta_j^2u_j^2 \right \}
    \end{equation}
Utilizing Young's inequality, given as $xy<\frac{x^2}{2\xi}+\frac{\xi y^2}{2}$, where $\xi$ represents an infinitesimal value, we obtain
\vspace{-0.1cm}
\begin{equation}
\label{X4}
    \small{\Delta V \leq \sum_{j=1}^{M}\left \{ -\beta_j(1-\frac{\xi_j}{2}-\frac{M}{2}\beta_j)u_j^2+\frac{\beta_j}{2\xi_j} \left [ \sum_{m \in N_i}(\Upsilon_m-\hat{\Upsilon}_j) \right ]^2  \right \}}
\end{equation}
With $N_j$ terms in $\sum_{m \in N_j}(\Upsilon_m-\hat{\Upsilon}_j)$ and using the sum of squares inequality, we get
\vspace{-0.1cm}
    \begin{equation}
    \label{X5}
        \left [ \sum_{m \in N_i}(\Upsilon_m-\hat{\Upsilon}_j) \right ]^2 \leq |M_j| \sum_{m \in N_j} (\Upsilon_m-\hat{\Upsilon}_j)^2
    \end{equation}
Substituting \eqref{X5} in \eqref{X4}, we obtain
\vspace{-0.1cm}
    \begin{equation}
    \label{X6}
        \small{\Delta V \leq \sum_{j=1}^{M}\left \{ -\beta_j(1-\frac{\xi_j}{2}-\frac{M}{2}\beta_j)u_j^2+\frac{\beta_j |N_j|}{2\xi_j} \sum_{m \in N_j} (\Upsilon_m-\hat{\Upsilon}_j)^2  \right \}}
    \end{equation}
Since the triggering instants in $j^{th}$ agent during data availability attacks are evaluated by
\vspace{-0.1cm}
\begin{equation}
\label{X7}
    u_j^2(k)=\gamma_i\hat{u_j^2}(k)
\end{equation}
\vspace{-0.1cm}
    \begin{equation}
    \label{X8}
        (\Upsilon_m(k)-\hat{\Upsilon_j(k)})^2\leq\frac{\sum_{j=1}^{M}\frac{\gamma_j\beta_j}{N_j}(1-\frac{\xi_j}{2}-\frac{M}{2}\beta_j)\hat{u}_j^2}{\sum_{j=1}^{M}\frac{\beta_j N_j}{2\xi_j}}
    \end{equation}
Adding and subtracting $\gamma_j\beta_j(1-\frac{\xi_j}{2}-\frac{M}{2}\beta_j)\hat{u}_j^2$ in \eqref{X6},
\vspace{-0.1cm}
    \begin{equation}
    \label{X9}
    \begin{aligned}
        \Delta V \leq & -\sum_{j=1}^{M}\beta_j(1-\frac{\xi_j}{2}-\frac{M}{2}\beta_j)(u_j^2-\gamma_j\hat{u_j}^2)+ \\ & \sum_{j=1}^{M}\left [\frac{\beta_j |N_j|}{2\xi_j} \left ( 1-\frac{\xi_j}{2}-\frac{M\beta_j}{2} \right )\hat{u}_j^2\sum_{m \in N_j} (\Upsilon_m-\hat{\Upsilon}_j)^2   \right ]
    \end{aligned}
    \end{equation}
\textit{Theorem 1:} $\Delta V(\Upsilon) \leq 0$ is guaranteed for all $k$ using \cref{X7,X8,X9} for any $j \in M$ and $m \in N_j$. The only scenario where $\Delta V = 0$ can happen when
\vspace{-0.1cm}
\begin{equation}
\label{X10}
    \left\{\begin{matrix}
u_j=\hat{u}_j=0 \hspace{0.5cm}\forall j \in M\\ 
\Upsilon_j=\hat{\Upsilon}_j=0 \hspace{0.5cm}\forall m \in N_j
\end{matrix}\right.
\end{equation}
\textit{Theorem 2:} Using \eqref{X10}, it is proved that $\hat{\Upsilon}(k)$ is asymptotically stable and converges to the semantic sampling signals.

%lbrt
\section{Performance Evaluation}
A real-time simulation testbed setup \cite{R30}, used to test the feasibility of the proposed delay-aware semantic sampling scheme is shown in Fig. \ref{fig:Proposed}(b). It comprises of OP-5700 
\vspace{-0.2cm}
\begin{table}[h!]
  \centering
  \caption{Test System Parameters}
    \begin{tabular}{|c|c|c|}
    \hline
    \multicolumn{3}{|c|}{\textbf{Parameters for DERs}} \\
    \hline
    Parameter & Symbol & Rating \\
    \hline
    %& &\\
    Power rating & $\mathrm{P}$ & 32 kW \\
    %\hline
    Nominal V and $\omega$ & $\mathrm{V_{nom}}$, $\mathrm{\omega_{nom}}$ & 220$\sqrt2$ V, 314.15 rad/s \\
    %\hline
    Filter parameters & $\mathrm{L^f}$, $\mathrm{r^f}$, $\mathrm{C^f}$ & 3 mH, 1 m$\Omega$, 12.1 mF \\
    %\hline
    Output impedance & $\mathrm{L^o}$, $\mathrm{r^o}$ & 1 mH, 0.121 $\Omega$ \\
    %\hline
   P droop coefficient & $\mathrm{m^{p}}$ & 9.4$\times$10$^{-5}$ rad/(W.s) \\
    Q droop coefficient & $\mathrm{n^{q}}$ & 1.3$\times$10$^{-3}$V/VAr \\
    Proportional gain (CC, VC) & $\mathrm{K_{p}^{i}}$, $\mathrm{K_{p}^{V}}$ & 0.2, 50 \\
    \rule{0pt}{3ex}
    Integral gain (CC, VC) & $\mathrm{K_{i}^{i}}$, $\mathrm{K_{i}^{V}}$ & 1, 100 \\
    %& &\\
    \hline
    \multicolumn{3}{|c|}{\textbf{Secondary control (SC) parameters}} \\
    \hline
     Communication weight & $\mathrm{a}_{jm}$ & 1 \\
    %\hline
    Convergence parameter & $\mathrm{g}_j$ & 1 \\
    %\rule{0pt}{3ex} 
    %& &\\
    Proportional gain & $\mathrm{K_{p}^{S\omega}}$, $\mathrm{K_{p}^{SV}}$ & 0.1, 0.1 \\
    \rule{0pt}{3ex}  
    Integral gain & $\mathrm{K_{i}^{S\omega}}$, $\mathrm{K_{i}^{SV}}$ & 42, 1.5 \\
    %& &\\
    \hline
    \multicolumn{3}{|c|}{\textbf{Network and load parameters}} \\
    \multicolumn{3}{|c|}{\textbf{of modified IEEE 37-bus AC distribution system} (\cite{R39})} \\
    \hline
    Alpha & $\alpha$ & 0.3 \\
    \hline
    \end{tabular}%
  \label{tab:Parameters}%
\end{table} (real-time simulator), which is integrated with HYPERSIM software (on the host PC) to model the required test system. The PC and OP-5700 simulator are seamlessly linked via an Ethernet interface, facilitating the establishment of IEC 61850 sampled values protocol for efficient communication and data exchange. The standard IEEE 37-bus system is modified by adding seven inverters at buses B 15, B 18, B 22, B 24, B 29, B 33, and B 34 as shown in Fig. \ref{fig:DER}(a). This modified test system is considered to validate the proposed delay-aware semantic sampling approach. The design and control parameters of DERs is provided in Table \ref{tab:Parameters}. The evaluation of this proposed scheme for various test conditions of latency attacks, data dropouts, TSAs is presented further.

\subsection{System under latency attacks}
\begin{figure}[h!]
        \centering
	    \includegraphics[clip, trim=0.5cm 5.3cm 8cm 0.5cm, width=0.96\linewidth]{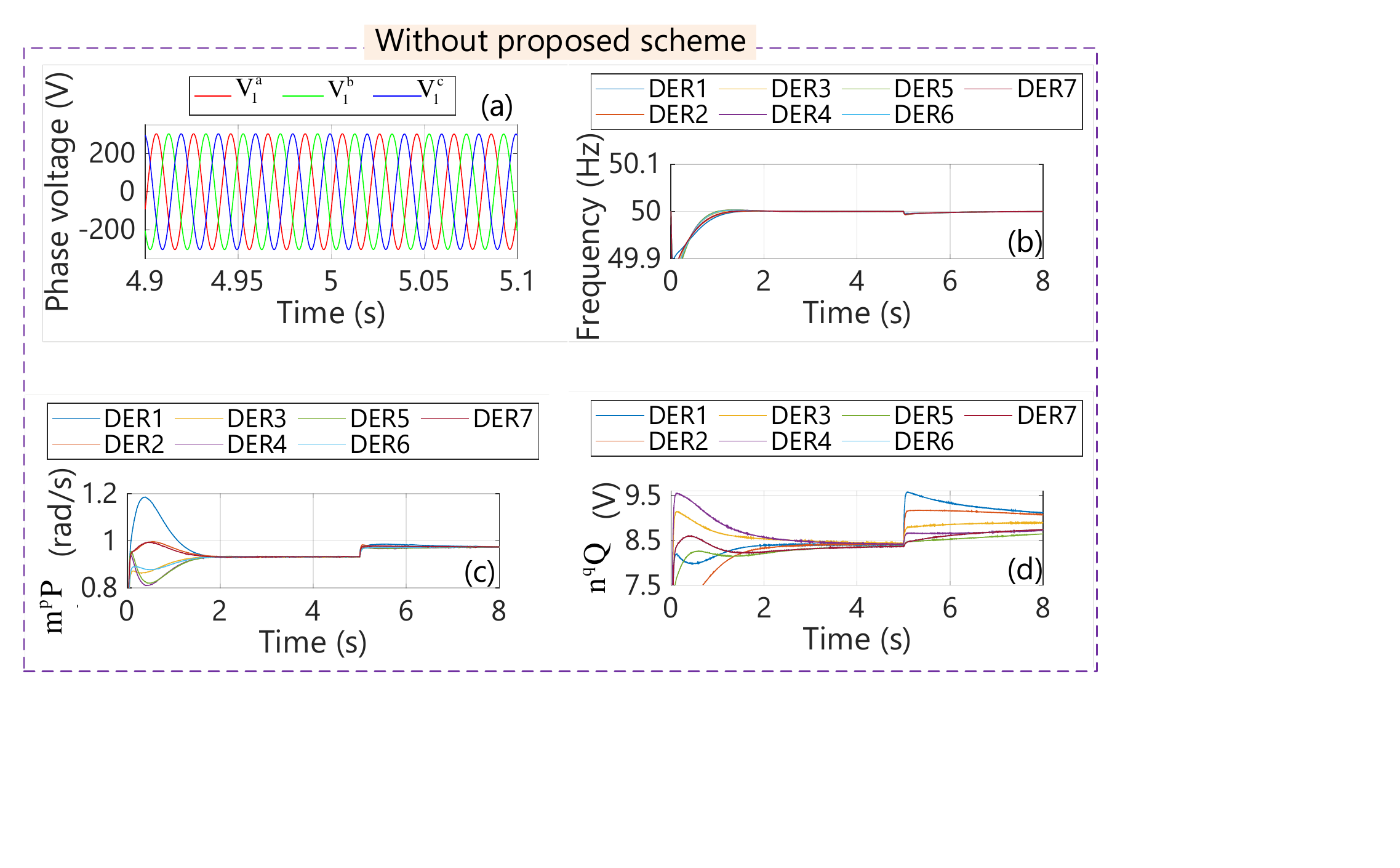}
	    \caption{Time-domain signals during latency attack ($\tau_m$=0.05 s), without the proposed scheme for \textbf{(a)} phase voltage of DER 1; \textbf{(b)} frequency; \textbf{(c)} active power sharing; and \textbf{(d)} reactive power sharing for all DERs.}
	    \label{fig:L_Prob}
\end{figure}

A latency attack was carried out on the considered system with the time delay, $\tau_m$ = 0.05 s. It was then followed by load variation at 5 s. Although the voltage remains within acceptable bounds, as shown in Fig. \ref{fig:L_Prob}(a), but the SC objectives are not accomplished. It can be observed from the time-domain plots of frequency, active and reactive power sharing (as in Fig. \ref{fig:L_Prob}(b), \ref{fig:L_Prob}(c), and \ref{fig:L_Prob}(d), respectively) that the consensus convergence time is increased due to delay.  \begin{figure}[h!]
        \centering
	    \includegraphics[clip, trim=0.5cm 5.3cm 8cm 0.5cm, width=0.96\linewidth]{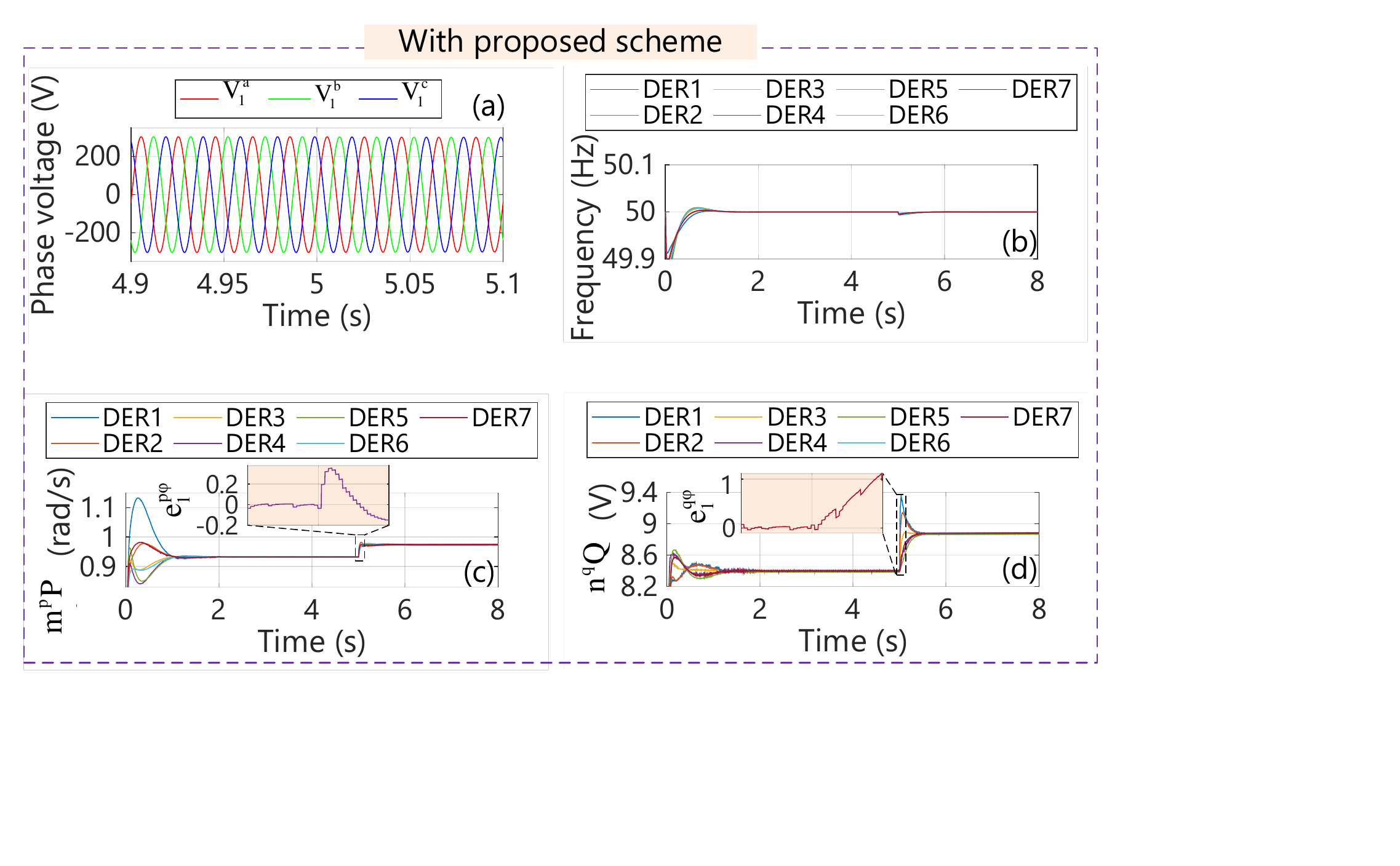}
	    \caption{Time-domain signals during latency attack ($\tau_m$=0.05 s), with the proposed scheme for \textbf{(a)} phase voltage of DER 1; \textbf{(b)} frequency; \textbf{(c)} active power sharing; and \textbf{(d)} reactive power sharing for all DERs.}
	    \label{fig:L_sol}
\end{figure} 

However, with the inclusion of the proposed scheme, resulting reconstructed signals ($\mathrm{e}_{j}^{\mathrm{p}\varphi}$ and $\mathrm{e}_{j}^{\mathrm{q}\varphi}$) as shown in Fig. \ref{fig:L_sol}(c) and \ref{fig:L_sol}(d) compensates for delay. It can be seen from time-domain plots of frequency, active and reactive power sharing (as in Fig. \ref{fig:L_sol}(b), \ref{fig:L_sol}(c), and \ref{fig:L_sol}(d), respectively), that convergence is much faster, with steady-state settling time within 0.45 s.  

\subsection{System under latency attacks and data dropouts}
\begin{figure}[h!]
        \centering
	   \includegraphics[clip, trim=0.5cm 6.5cm 8cm 0.5cm, width=1\linewidth]{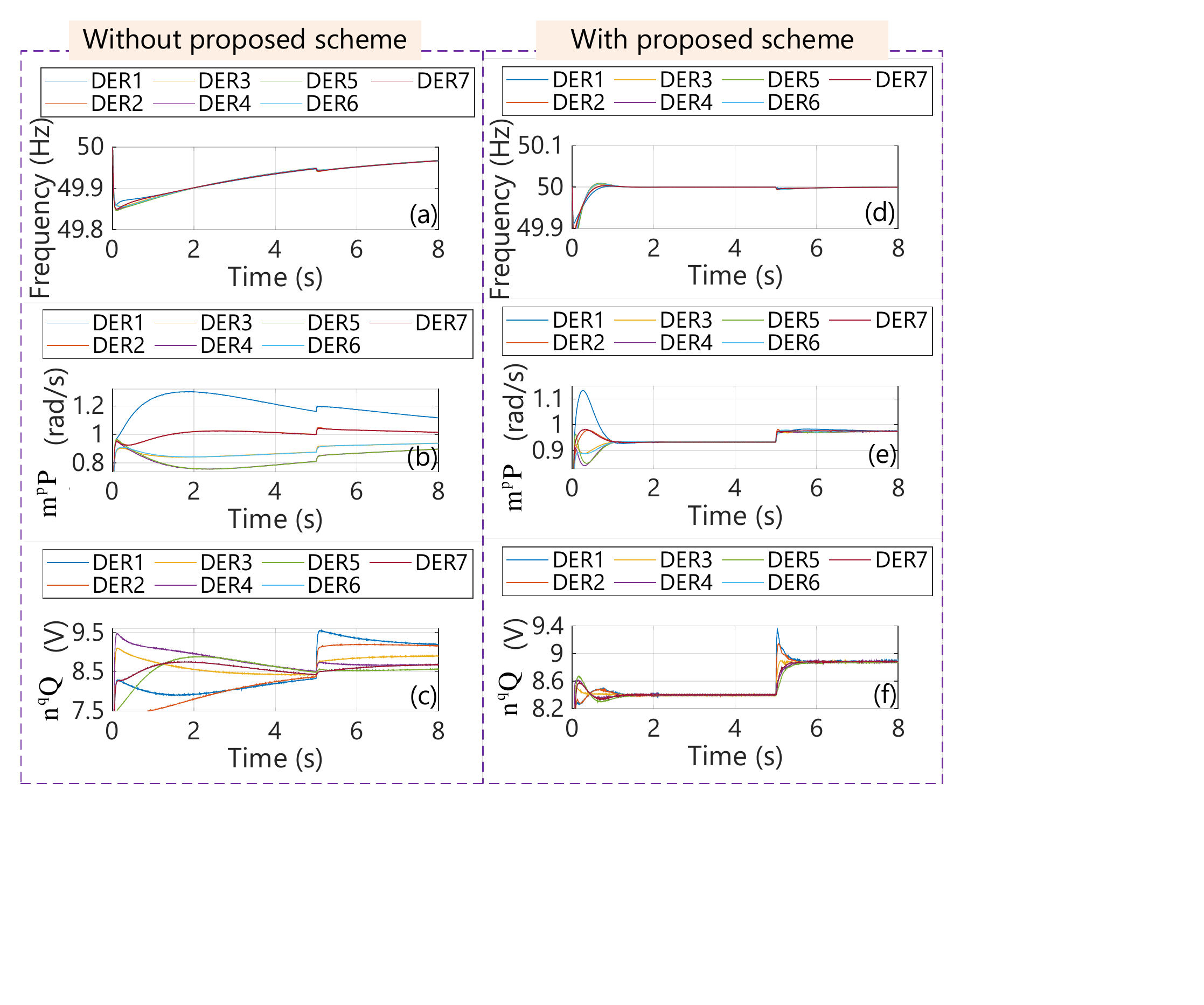}
	    \caption{The time-domain signals during latency attack ($\tau_m$=0.05 s) and 10\% data dropout for \textbf{(a)} frequency; \textbf{(b)} active power sharing; \textbf{(c)} reactive power sharing, without the proposed scheme are shown. Further, the time-domain signals for \textbf{(d)} frequency; \textbf{(e)} active power sharing; \textbf{(f)} reactive power sharing, with the proposed scheme are shown.}
	    \label{fig:LD}
\end{figure}

Considering a latency attack ($\tau_m$ = 0.05 s) with 10\% data dropout and load variation at 5 s, it can seen in Fig. \ref{fig:LD}(a), \ref{fig:LD}(b), and \ref{fig:LD}(c), that time required to attain SC objectives is further increased as compared to initiation of only latency attack as in case A. Further, with proposed scheme SC objectives are attained at much faster rate as seen in Fig. \ref{fig:LD}(d), \ref{fig:LD}(e), and \ref{fig:LD}(f) for frequency, active and reactive power sharing, respectively. This can be attributed to reconstructed signal from the local controller that drives the control process during such attacks.

\subsection{System under TSAs}
\begin{figure}[h!]
        \centering
	    \includegraphics[clip, trim=0.5cm 7cm 8cm 0.5cm, width=1\linewidth]{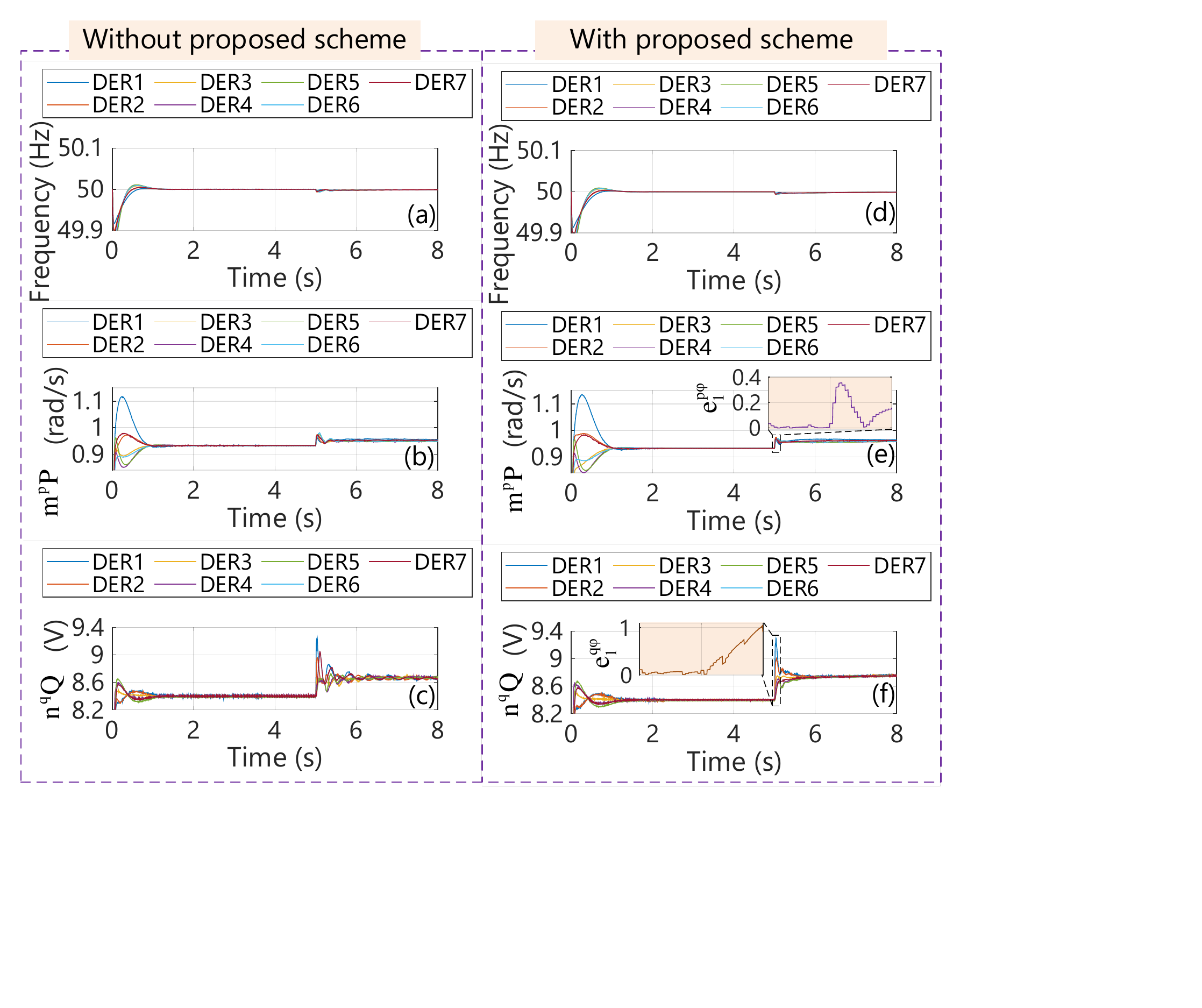}
	    \caption{The time-domain signals during TSA for \textbf{(a)} frequency; \textbf{(b)} active power sharing; \textbf{(c)} reactive power sharing, without the proposed scheme are shown. Further, the time-domain signals for \textbf{(d)} frequency; \textbf{(e)} active power sharing; \textbf{(f)} reactive power sharing, with the proposed scheme are shown.}
	    \label{fig:T}
\end{figure}

Similarly, the time-domain simulation for the system under TSA attack (considering load variation at 5 s) without the proposed scheme is shown in Fig. \ref{fig:T}(a), \ref{fig:T}(b), and \ref{fig:T}(c).  It can be observed that with the deployment of the proposed scheme, the dynamic performance is increased because of the reconstructed signals ($\mathrm{e}_{j}^{\mathrm{p}\varphi}$ and $\mathrm{e}_{j}^{\mathrm{q}\varphi}$), as shown in Fig. \ref{fig:T}(e) and \ref{fig:T}(f). It can be seen in Fig. \ref{fig:T}(d), \ref{fig:T}(e), and \ref{fig:T}(f), that the convergence time to attain SC of frequency restoration, proportional active and reactive power sharing decreases. 

\subsection{Cyber graph variations}
\begin{figure}[h!]
        \centering
	    \includegraphics[clip, trim=0.5cm 7cm 8cm 0.5cm, width=1\linewidth]{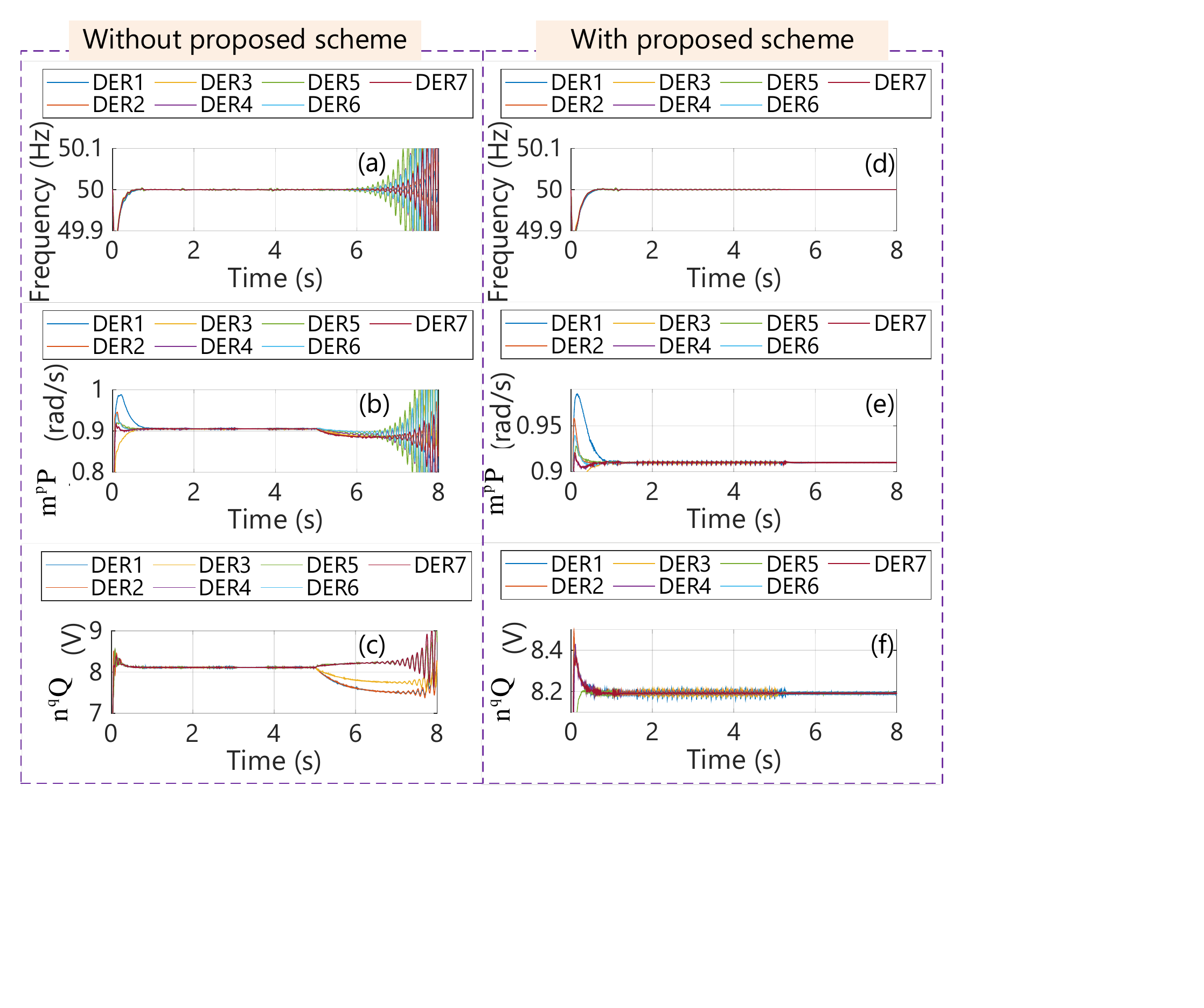}
	    \caption{The time-domain signals during latency attack ($\tau_m$=0.05 s) with cyber graph variations for \textbf{(a)} frequency; \textbf{(b)} active power sharing; \textbf{(c)} reactive power sharing, without the proposed scheme are shown. Further, the time-domain signals for \textbf{(d)} frequency; \textbf{(e)} active power sharing; \textbf{(f)} reactive power sharing, with the proposed scheme are shown.}
	    \label{fig:TP}
\end{figure}

The system is tested for two cyber graphs (G) under latency attack ($\tau_m$ = 0.05 s). These graphs are G1 and G2 representing fully-connected graph and ring-connected graph, respectively. Initially, the system is connected in G1 configuration and later switched to G2 configuration at 5 s. It can be observed from \ref{fig:TP}(a), \ref{fig:TP}(b) and Fig. \ref{fig:TP}(c), that the system tends towards instability under latency attack  followed with dynamic cyber graph variations. This instability can be attributed to both the resulting sparse network and the additional delay to the signals, due to which the agents were not able to update their states continuously thereby slowing down the convergence. However, delay-aware semantic sampling actively synchronizes the error signals at primary and secondary controllers to generate reconstructed signals, thereby making this proposed scheme robust to dynamic cyber graph variations along with the latency attack as shown in Fig. \ref{fig:TP}(d), \ref{fig:TP}(e) and Fig. \ref{fig:TP}(f).

\begin{figure}[h!]
        \centering
	    \includegraphics[clip, trim=0.5cm 3cm 7cm 0.5cm, width=1\linewidth]{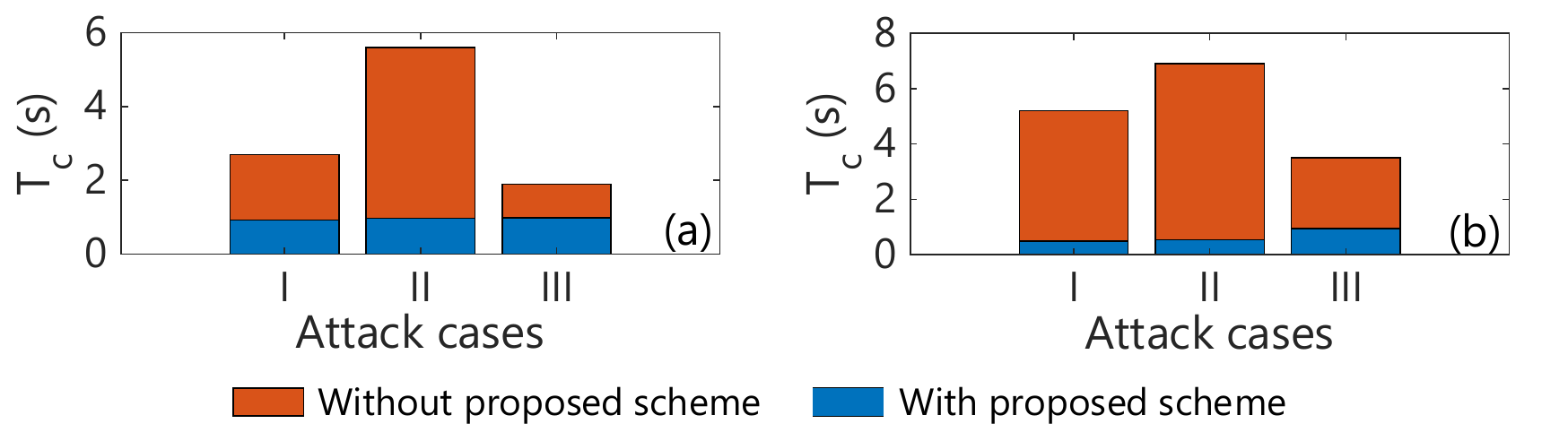}
	    \caption{Plots of time of convergence ($\mathrm{T_c}$) for \textbf{(a)} O1; and \textbf{(b)} O2, without and with the deployment of the proposed scheme.}
	    \label{fig:TC}
    \end{figure}

Let us consider that different attacks are represented as: a) I: latency attack; b) II: latency attack and data dropout; and c) III: TSA. The convergence time ($\mathrm{T_c}$) plots for the system under these attacks are depicted in Fig. \ref{fig:TC}.  Let O1 and O2 are defined as the objectives of SC. Here, O1 is attaining frequency restoration and proportional active power sharing; and O2 is attaining proportional reactive power sharing. It is evident from Fig. \ref{fig:TC}(a) and \ref{fig:TC}(b), that cases without the proposed delay-aware semantic sampling scheme, particularly those subjected to latency attacks with $\tau_m$=0.05 s and 10\% data dropout, exhibit longer convergence times compared to instances featuring only latency attacks with $\tau_m$=0.05 s. In contrast, the deployment of our proposed delay-aware semantic sampling scheme substantially reduces convergence times across all attack scenarios (as shown in Fig. \ref{fig:TC}(a) and \ref{fig:TC}(b)), thereby enhancing overall system performance. 
    \begin{table*}[b!]
\caption{Comparative Analysis of the Proposed Delay-Aware Semantics Scheme for PES.}
\centering
\def\arraystretch{0.6}
\label{table:comp}
\begin{tabular}{|>{\color{black}}p{0.04\linewidth}|>{\color{black}}p{0.2\linewidth}|>{\color{black}}p{0.15\linewidth}|>{\color{black}}p{0.15\linewidth}|>{\color{black}}p{0.15\linewidth}|>{\color{black}}p{0.15\linewidth}|}
\hline
Sl. no. & Features & \cite{R23} & \cite{R25} & \cite{R26} & \textbf{Proposed scheme}\\
\hline
%&&&&&\\
1. & Computational complexity & Medium & Medium & High & Low\\
%&&&&\\
2. & Distributed concept & \xmark & \xmark & \xmark & \cmark\\
%&&&&\\
3. & Resilient to latency attacks & \cmark & \cmark & Not tested & \cmark \\
%&&&&\\
4. & Resilient to TSAs & Not tested & Not tested & Only detection & \cmark \\
%&&&&\\
5. & Resilient to data dropouts & Not tested & Not tested & Not tested & \cmark \\
%&&&&\\
6. & Robust to loading variations & \cmark & \cmark & \cmark & \cmark \\
%&&&&\\
7. & Model-agnostic & \xmark & \xmark & \cmark & \cmark \\
%&&&&\\
8. & Supports dynamic cyber graphs & \cmark & Not tested & Not tested & \cmark \\
%&&&&\\
\hline
\end{tabular}
\end{table*}
    Additionally, the steady-state error is assessed for the aforementioned attack scenarios to achieve SC objectives O1 and O2. Specifically, we define the absolute value of the steady-state error as \begin{figure}[h!]
        \centering
	    \includegraphics[clip, trim=0.5cm 3cm 7cm 0.5cm, width=0.9\linewidth]{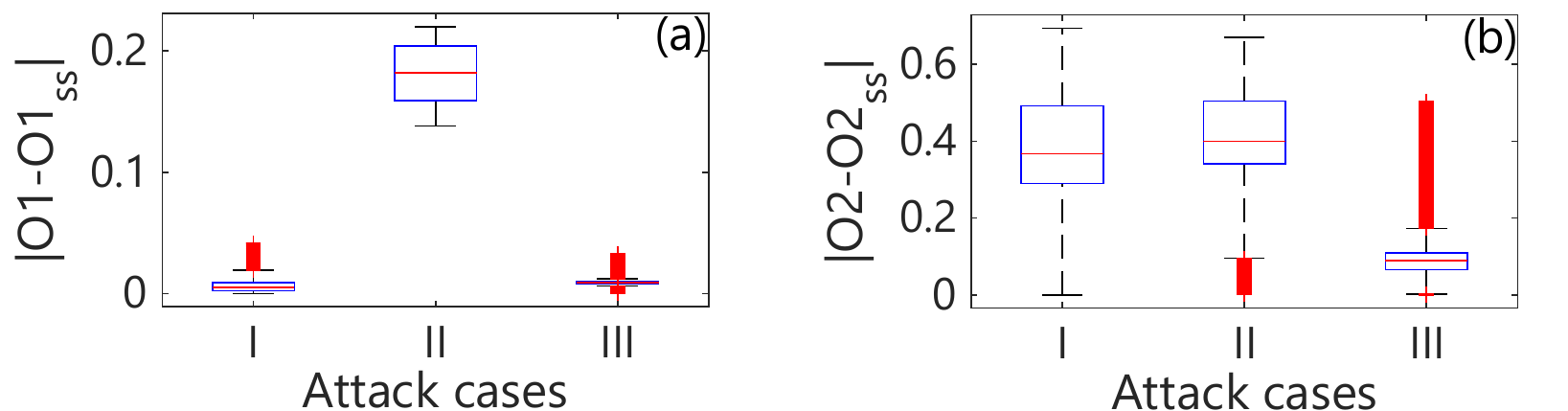}
	    \caption{Plots of steady-state error with different attack cases for \textbf{(a)} O1; and \textbf{(b)} O2, without the proposed scheme.}
	    \label{fig:SSE_P}
    \end{figure} \begin{figure}[h!]
        \centering
	    \includegraphics[clip, trim=0.5cm 3cm 7cm 0.5cm, width=0.9\linewidth]{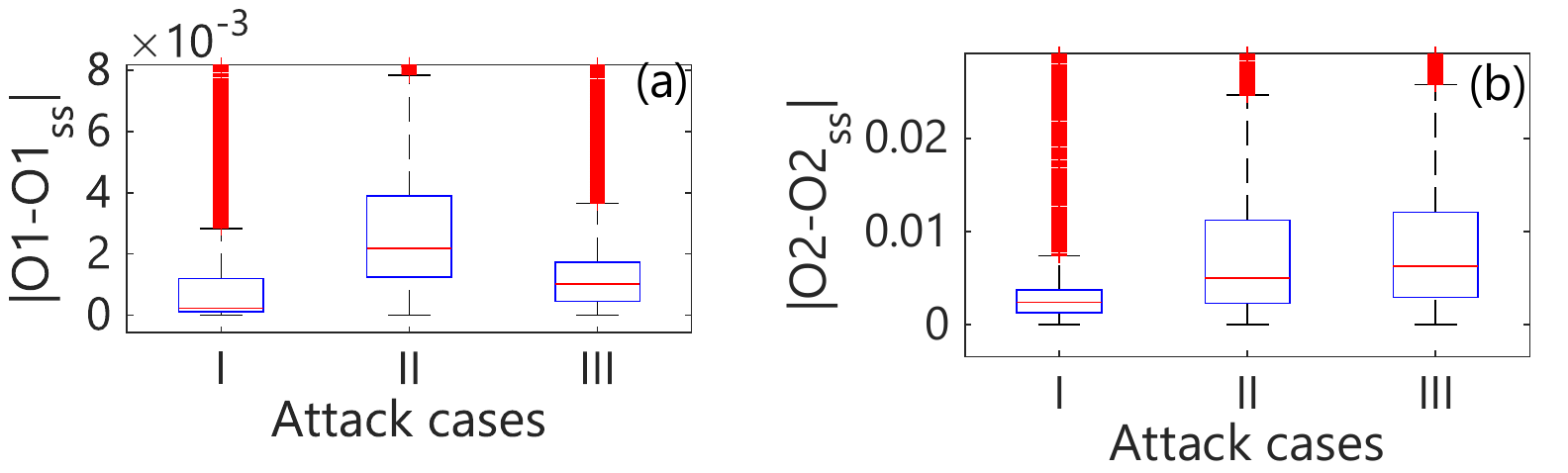}
	    \caption{Plots of steady-state error with different attack cases for \textbf{(a)} O1; and \textbf{(b)} O2, with the proposed scheme.}
	    \label{fig:SSE_S}
    \end{figure} $|\mathrm{O1}-\mathrm{O1_{ss}}|$ and $|\mathrm{O2}-\mathrm{O2_{ss}}|$ for O1 and O2, respectively. Here $\mathrm{O1}$ and $\mathrm{O2}$ represent the instantaneous values, and $\mathrm{O1_{ss}}$ and $\mathrm{O2_{ss}}$ denote the steady-state values, according to SC objectives. Fig. \ref{fig:SSE_P}(a) and \ref{fig:SSE_P}(b) illustrate the steady-state error for achieving O1 and O2, respectively, in the system without the proposed scheme. Conversely, Fig. \ref{fig:SSE_S}(a) and \ref{fig:SSE_S}(b) reveal significantly reduced steady-state errors to attain O1 and O2 with the deployment of our proposed scheme, thereby improving system performance.
    
Further, the various error signals for latency attack ($\tau_m$=0.05 s) with cyber graph variations, \begin{figure}[h!]
        \centering
	    \includegraphics[clip, trim=0.5cm 7cm 8.5cm 0.5cm, width=1\linewidth]{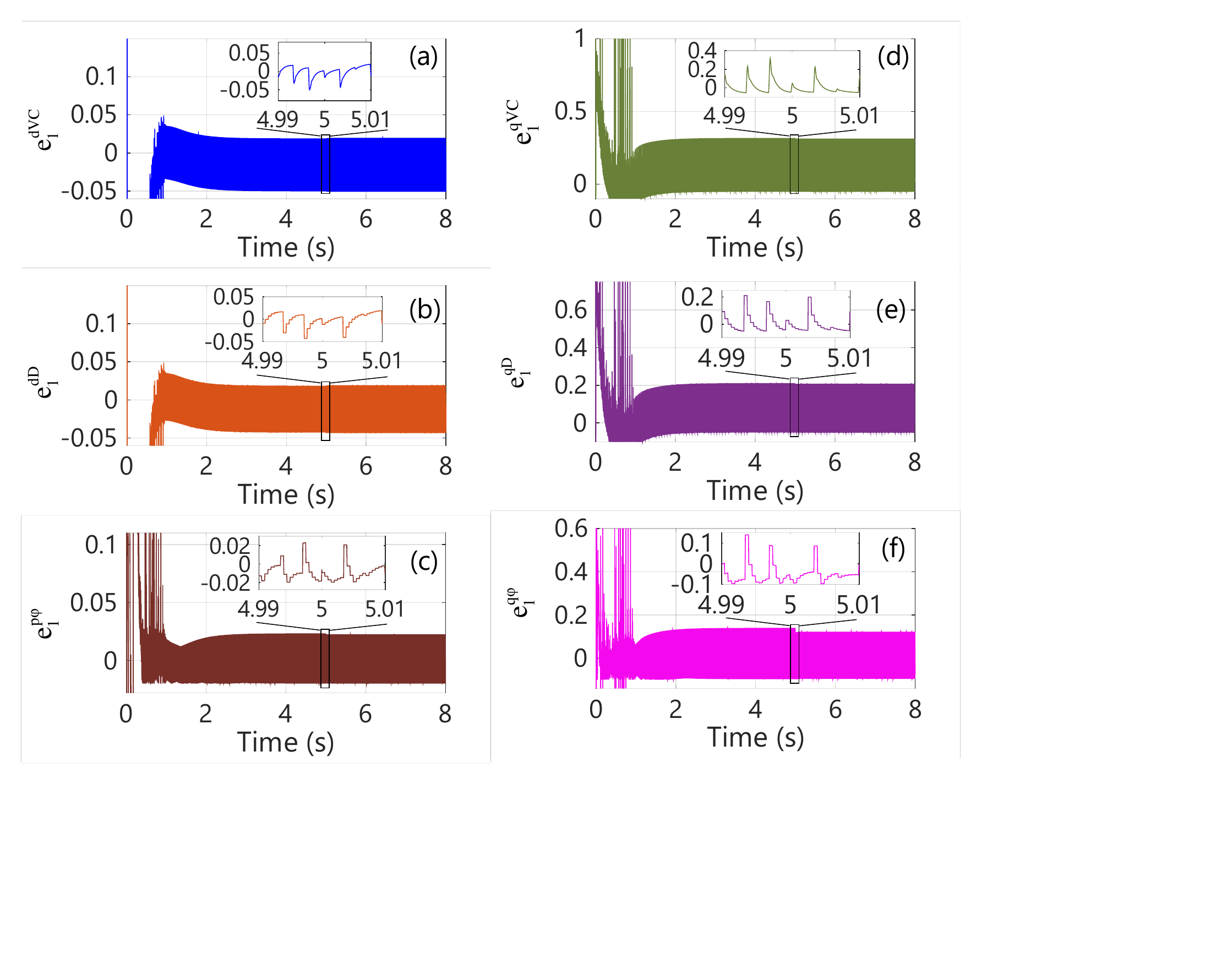}
	    \caption{The time-domain signals during latency attack ($\tau_m$=0.05 s) with cyber graph variations for \textbf{(a)} $d$-axis error signals input to VC, $\mathrm{e}_1^{\mathrm{dVC}}$; \textbf{(b)} process-aware sparsely sampled $d$-axis error signals input to VC, $\mathrm{e}_1^\mathrm{dD}$; \textbf{(c)} reconstructed signal input to SC, $\mathrm{e}_{1}^{\mathrm{p}\varphi}$; \textbf{(d)} $q$-axis error signals input to VC, $\mathrm{e}_1^{\mathrm{qVC}}$; \textbf{(e)} process-aware sparsely sampled $q$-axis error signals input to VC, $\mathrm{e}_1^\mathrm{qD}$; and \textbf{(f)} reconstructed signal input to SC, $\mathrm{e}_{1}^{\mathrm{q}\varphi}$.}
	    \label{fig:Error} 
\end{figure} at DER 1 are investigated. The error signals provided by VC ($\mathrm{e}_1^{\mathrm{dVC}}$, $\mathrm{e}_1^{\mathrm{qVC}}$) are shown in Fig. \ref{fig:Error}(a) and Fig. \ref{fig:Error}(d), respectively. The process-aware sparse sampling  of these signals are carried out as shown in Fig. \ref{fig:Error}(b) and Fig. \ref{fig:Error}(e), respectively. The resulting reconstructed signals which are fed to the SC for delay compensation are shown in Fig. \ref{fig:Error}(c) and Fig. \ref{fig:Error}(f), respectively.
 
\begin{figure}[h]
        \centering
	    \includegraphics[clip, trim=0.5cm 3.2cm 9cm 0.5cm, width=1\linewidth]{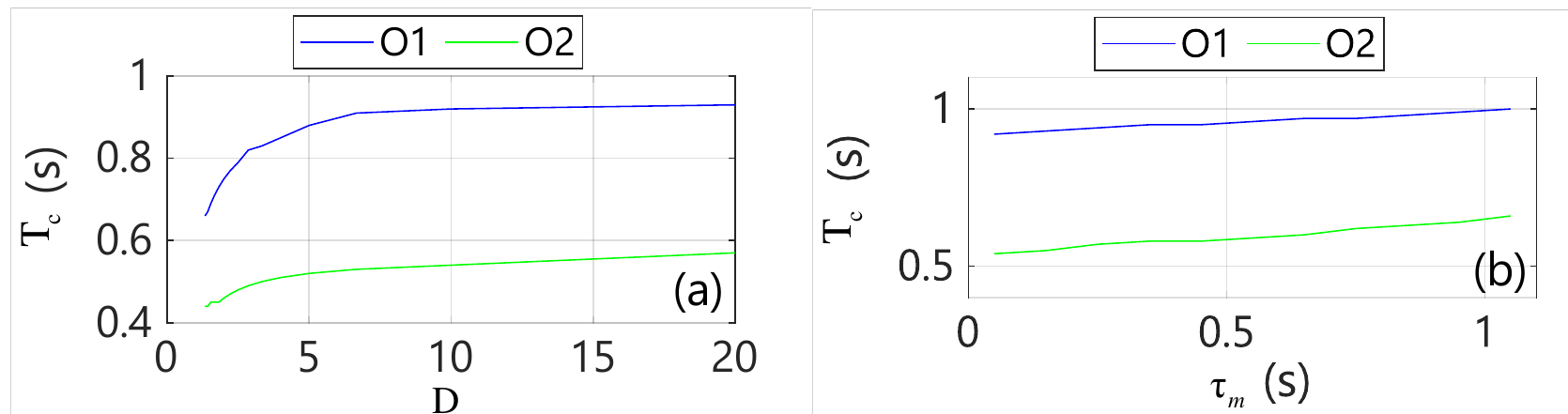}
	    \caption{Plots for variation in $\mathrm{T_c}$ for attaining O1 and O2 with variations in \textbf{(a)} downsampling factor (D); and \textbf{(b)} amount of delay ($\tau_m$).}
	    \label{fig:Tc} 
\end{figure}

The system's performance was evaluated by varying the downsampling factor ($\mathrm{D}$), as shown in Fig. \ref{fig:Tc}(a).  The results indicate a non-linear relationship between $\mathrm{D}$ and $\mathrm{T_c}$ to achieve the SC objectives. As expected, $\mathrm{T_c}$ increases with increasing $\mathrm{D}$. Therefore, selecting the appropriate value of $\mathrm{D}$ involves a trade-off between low-cost operation versus fast efficient performance. In this study, a value of 10 was chosen for $\mathrm{D}$, which offers a good balance of low-cost and fast efficient operation. The proposed delay-aware semantic sampling scheme was also tested for system performance under different levels of delay ($\tau_m$), as shown in \ref{fig:Tc}(b), demonstrating its robustness to large random delayed-measurements.  

\begin{figure}[h!]
        \centering
	   \includegraphics[clip, trim=0.5cm 4.4cm 4.5cm 0.5cm, width=1\linewidth]{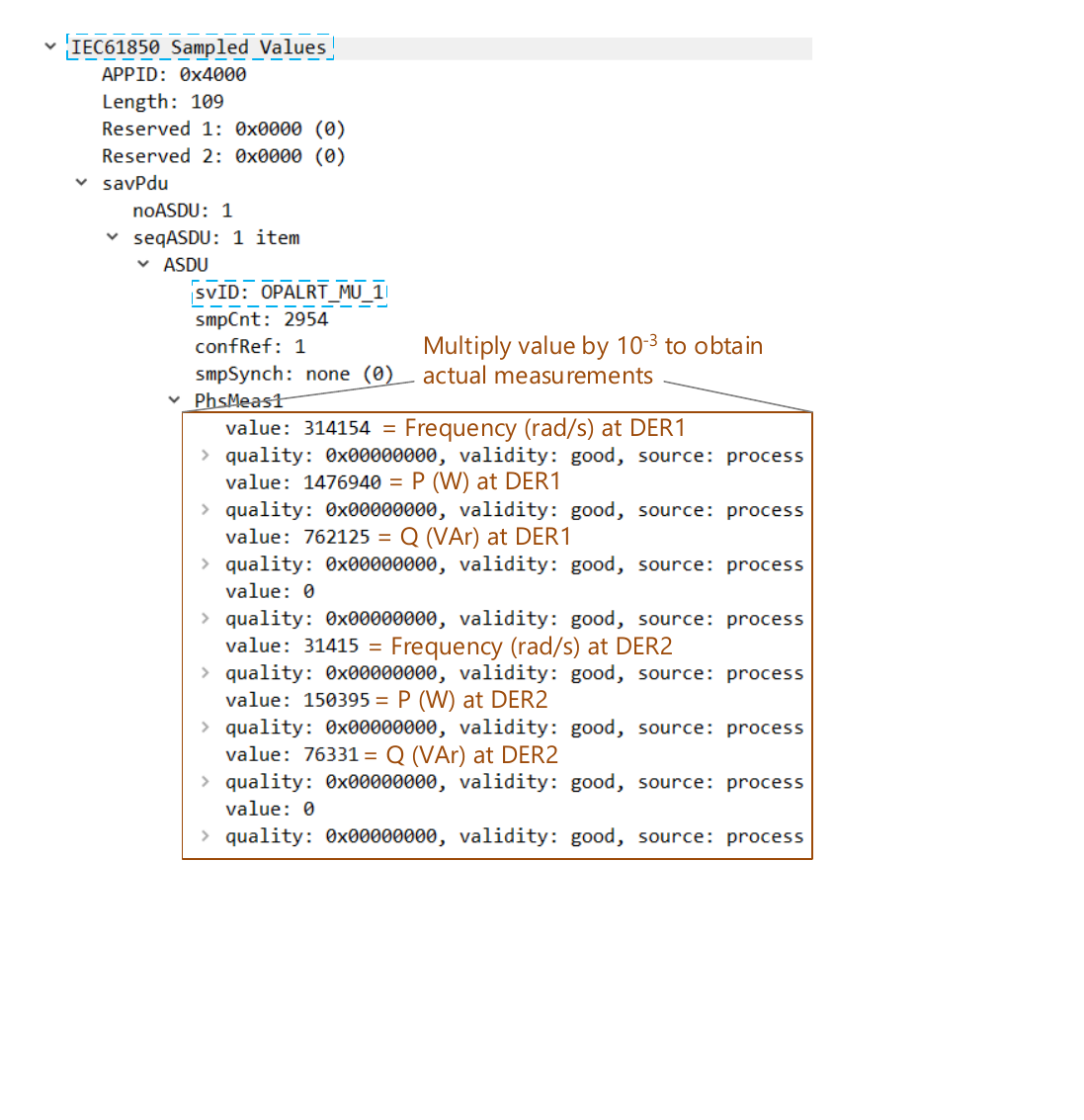}
	    \caption{Details of IEC 61850 sampled values packet comprising of svID, values of communicated signals etc., obtained from \textit{wireshark} application.}
	    \label{fig:Packet}
\end{figure}

Additionally, the cyber layer of a PES (comprising of two DERs) was developed over IEC 61850 sampled values protocol through Ethernet interface. This communication model is based on a publish-subscribe architecture \cite{R30}. \begin{figure}[h!]
        \centering
	    \includegraphics[clip, trim=0.5cm 5.8cm 8cm 0.5cm, width=1\linewidth]{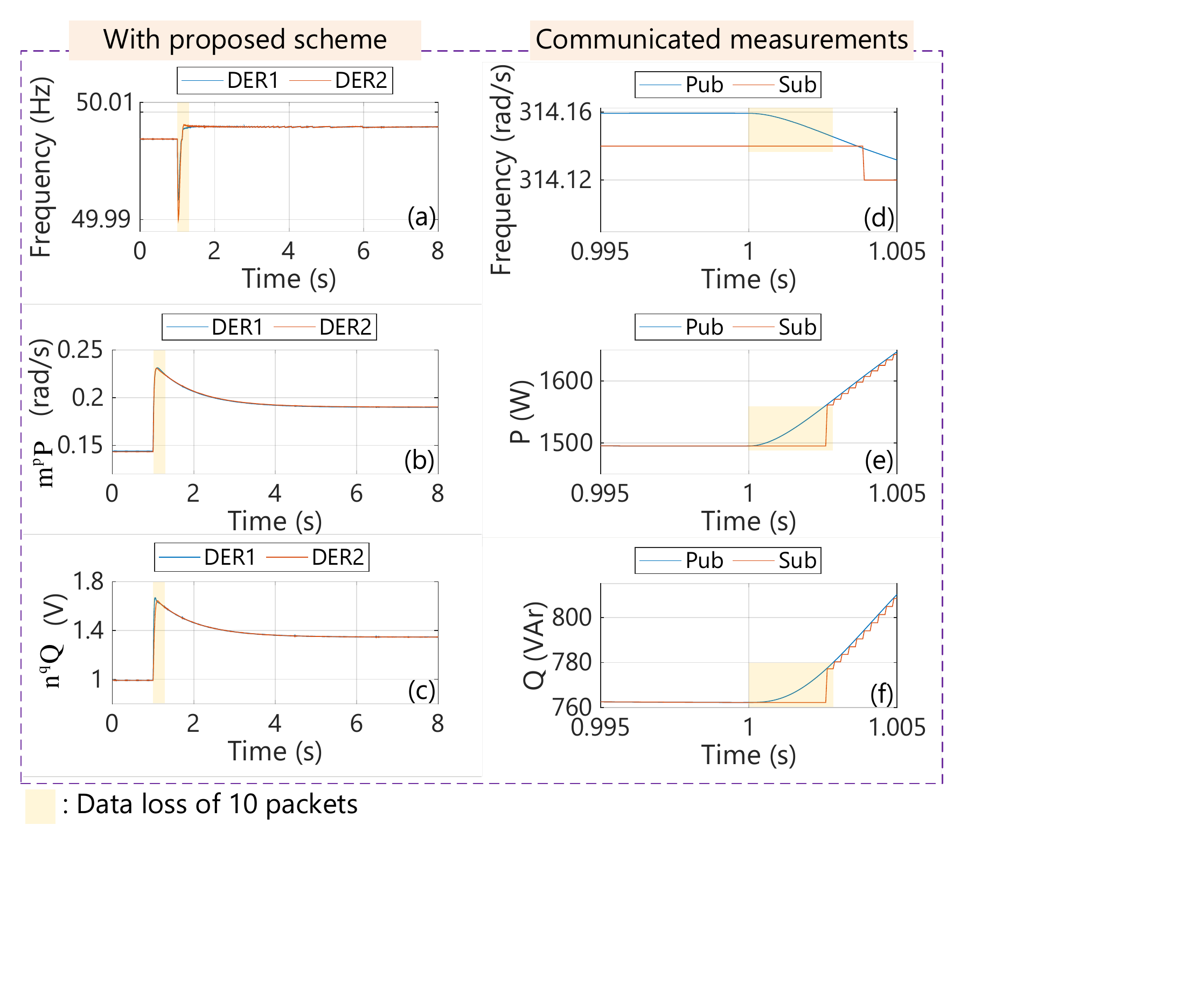}
	    \caption{The time-domain signals during data loss of 10 packets for \textbf{(a)} frequency; \textbf{(b)} active power sharing; and \textbf{(c)} reactive power sharing, with the proposed scheme are shown. Further, the time-domain signals showing data loss of 10 packets in the communicated signals i.e, \textbf{(d)} frequency; \textbf{(e)} active power; and \textbf{(f)} reactive power, over IEC 61850 sampled values protocol are presented.}
	    \label{fig:SV_DoS}
\end{figure} The data loss attack (of 10 packets) at 1 s (followed by load change at 1 s), was further tested on this system with the deployment of the proposed scheme. It can be observed from Fig. \ref{fig:SV_DoS}(a), \ref{fig:SV_DoS}(b) and \ref{fig:SV_DoS}(c)  that all objectives of SC are achieved. Moreover, the signals being published over the established protocol are indicated as ``\textit{Pub}" and the signals being subscribed are indicated as ``\textit{Sub}", as shown in Fig. \ref{fig:SV_DoS}(d), \ref{fig:SV_DoS}(e) and \ref{fig:SV_DoS}(f). Whenever data loss occurs, the subscribers hold on to the last received sample until the next packet arrives as shown in Fig. \ref{fig:SV_DoS}(d), \ref{fig:SV_DoS}(e) and \ref{fig:SV_DoS}(f). The details of the packet over IEC 61850 sampled values protocol are mentioned in Fig. \ref{fig:Packet}.

A comparative evaluation of our delay-aware semantic sampling scheme against existing methodologies is presented in Table \ref{table:comp}. The proposed scheme in this work, distinguishes itself as a computationally efficient solution, incorporating distributed concept that enhance its resilience against data availability attacks. Moreover, it demonstrates notable robustness when confronted with load variations, highlighting its practical adaptability. The model-agnostic nature of this approach further streamlines its implementation. Additionally, it supports dynamic cyber graphs, underscoring its practicality and flexibility. As a result, the delay-aware nodal semantic intelligence presented by our approach emerges as a highly promising and commercially viable solution, well-poised to address the intricate challenges within the realm of PES.

\section{Conclusions and Future Scope of Work}
In the landscape of PES, data availability challenges underscore the critical need for an innovative approach to mitigate the impact of random communication delays. Motivated by this imperative, our proposed delay-aware semantic methodology harnesses the inherent dynamics of the inner control loops within DERs to generate localized delay compensation signals. This approach not only yields robust performance and precise predictions by transmitting only the significant information but also obviates the need for intricate models and training that often accompany in prevailing methods. Real-time simulations on the OPAL-RT platform convincingly affirm the efficacy of our approach. While this study addresses the immediate challenges, several others loom ahead, such as understanding the limits of maximum communication delay tolerance and scalability in larger, complex systems. In future investigations, system stability under semantic sampling concerning the maximum communication delay it can handle will be explored and scalability in more extensive systems will be assessed.

In the evolving domain of demand response, diverse resources, such as electric vehicles, adaptable residential loads, and energy storage systems, are ready for integration. However, real-time data exchange among them necessitates varied communication protocols, posing a challenge for semantic interoperability. Our upcoming research aims to overcome this hurdle by developing a semantic framework to predict, activate, and manage heterogeneous resources efficiently. This research direction, not only promises to advance the field but also address the complexities of the PES energy market. Furthermore, semantic-based quantum communication will be explored to enhance fault detection and localization, reducing response times and downtime during disturbances, fortifying system resiliency.\\
\ifCLASSOPTIONcaptionsoff
  \newpage
\fi

%\newpage

\end{document}